\documentclass[11pt,a4paper]{article}
\usepackage{jcappub}
\usepackage{graphicx}
\usepackage[utf8]{inputenc} 
\usepackage{amsmath}
\usepackage{amssymb}
\usepackage{mathtools,braket,blkarray}
\usepackage{float}
\usepackage{comment}
\usepackage{slashed}
\usepackage[normalem]{ulem}
\usepackage{orcidlink}
\usepackage{multirow}
\usepackage{natbib}
\usepackage{soul}
\usepackage[title]{appendix}

\def\gagg{g_{a \gamma \gamma}}

\def\bal#1\eal{\begin{align}#1\end{align}}
\def\bea#1\eea{\begin{eqnarray}#1\end{eqnarray}}
\def\beq#1\eeq{\begin{equation}#1\end{equation}}

\def\nn{\nonumber}

\title{Solar axions from nuclear transitions}

\author[a]{Tanmoy Kumar\orcidlink{0000-0001-9775-6645}}
\author[b]{and Newton Nath\orcidlink{0000-0002-0592-0020}}

\affiliation[a]{School of Physical Sciences, Indian Association for the Cultivation of Science, 2A \& 2B Raja S.C. Mullick Road, Jadavpur, Kolkata 700032, India}
\affiliation[b]{Department of Physics, Indian Institute of Technology (BHU), Varanasi 221005, India}

\emailAdd{kumartanmoy1998@gmail.com}
\emailAdd{nnath.phy@iitbhu.ac.in}

\abstract{
We investigate the possibility of detecting   14.4~keV  and 9.4~keV solar axions and axion-like particles that could be produced in the M1 nuclear transitions of $^{57}\mathrm{Fe}$ and $^{83}\mathrm{Kr}$, respectively. 
To do so, we used data from soft X-ray observations of the quiet Sun collected by the Solar X-ray Monitor (XSM) on board India’s Chandrayaan-2 lunar mission. 
We observe that although the effective axion-nucleon couplings for $^{83}\mathrm{Kr}$ and $^{57}\mathrm{Fe}$ differ only slightly, their fluxes differ by nearly three orders of magnitude. Consequently, the limit on $|g_{aN}^{\rm eff} \times g_{a\gamma\gamma}|$ and only $\gagg$ vs.\ $m_a$ provide {more than one order
of magnitude stronger} constraint for Fe than for Kr.
}


\begin{document}

\maketitle

\section{Introduction}

The \emph{strong CP problem}, a longstanding unanswered puzzle in the Standard Model (SM) of particle physics, arises from the fact that charge conjugation-parity (CP) symmetry is expected to be violated in quantum chromodynamics (QCD), yet no such violation has been observed experimentally. 
One of the most elegant solutions to this problem is the introduction of an anomalous global $U(1)$ symmetry, known as the Peccei-Quinn (PQ) symmetry, which is spontaneously broken. 
This mechanism gives rise to a pseudo-Nambu-Goldstone boson (pNGB) called the QCD axion~\cite{Peccei:1977hh,Wilczek:1977pj,Weinberg:1977ma}.
At present, the axion remains one of the best-motivated elementary particles beyond the SM (BSM). 
%
Furthermore, axions are also excellent dark matter candidates~\cite{Dine:1982ah,Abbott:1982af,Preskill:1982cy,Davis:1986xc,Kim:2008hd,Marsh:2015xka,Adams:2022pbo,Berti:2022rwn}. 
The theoretical framework admits a large variety of QCD axion models, each characterized by distinctive phenomenological features. For a comprehensive review, see Ref.~\cite{DiLuzio:2020wdo} and references therein.
Apart from QCD axions, a broader class of pseudoscalar bosons, called axion-like particles (ALPs), emerges in many BSM scenarios \cite{Jaeckel:2010ni}. {Hereafter}, we will use the term ``axions" to refer to both axions and ALPs.
Due to their compelling theoretical motivation, a massive effort is underway to search for these elusive axions,
~\cite{Giannotti:2017hny,Agrawal:2021dbo,DiLuzio:2021ysg}. This creates significant excitement for the possible discovery of such BSM particles in the next decade~\cite{Giannotti:2017law,Giannotti:2022euq}.
For recent reviews, see Refs.~\cite{Irastorza:2018dyq,DiVecchia:2019ejf,DiLuzio:2020wdo,Agrawal:2021dbo,Sikivie:2020zpn,Giannotti:2022euq,Arza:2026rsl} and the references therein.
For an overview of the latest constraints, see~\cite{AxionLimits}.

In particular, astrophysical objects such as stars, supernovae, quasars, etc., in lieu of their extreme temperatures and densities, act as efficient factories of these axions. Using the observation of such objects via highly sensitive telescopes across a wide range of frequencies, some of the most stringent constraints on {axion parameter space}
have been obtained~\cite{Vysotsky:1978dc,Raffelt:2006cw,Lella:2022uwi,Lella:2023bfb,Carenza:2024ehj,Lecce:2025dbz,Fiorillo:2025gnd,Candon:2025sdm, Haxton:2025xqz,Haxton:2026ura,Fiorillo:2026rgo}.
A particularly important target for these astrophysical axion searches is the Sun. 
In the dense solar core, where temperatures $\sim 1\,\mathrm{keV}$, axions with masses below a few keV can be efficiently produced via their coupling to photons 
through the Primakoff effect~\cite{Raffelt:1985nk,Raffelt:1987np} and photon-axion conversion in solar magnetic fields~\cite{Caputo:2020quz,OHare:2020wum,Guarini:2020hps}, as well as via their coupling to electrons 
through electron-nucleus scattering, electron-electron bremsstrahlung, and Compton processes~\cite{Redondo:2013wwa, Hoof:2021mld}.
%
The CERN Axion Solar Experiment (CAST)~\cite{CAST:2024eil} has searched for these solar axions and, in the absence of any signal, has placed a tight constraint on axions' coupling to photons ($g_{a\gamma\gamma}$), excluding 
{$ g_{a\gamma\gamma} > 5.8 \times 10^{-11}$ GeV$ ^{-1} $} at 95\% confidence level for axions with mass $m_a\lesssim 0.02$~eV.
A similar bound can be obtained from observations of horizontal branch (HB) stars in globular clusters~\cite{Ayala:2014pea,Straniero:2015nvc}. 
Recently, authors of~\cite{Dolan:2022kul} have
found a slightly stronger bound, specifically $\gagg\lesssim 0.34 \times 10^{-10}$~GeV$^{-1}$ for $m_a<1$~keV, by measuring the ratio of stars in the asymptotic giant branch and in the HB in globular clusters.

Besides these, axions also interact with nucleons such as neutrons and protons. This presents the possibility that
axions can be produced from the de-excitation of thermally excited nuclei of elements present in the core of the Sun.
The Sun, being a Population I star, has a significant abundance of heavy 
elements~\footnote{By heavy elements we mean elements heavier than hydrogen.}, 
which have either been synthesized through nuclear reactions occurring in the 
solar interior or have been inherited as remnants of the primordial molecular cloud from 
which the solar system formed. Among these, odd-$A$ nuclei possessing low-lying 
excited states with excitation energies of a few keV can be thermally excited 
via the absorption of photons in the hot solar plasma. The subsequent 
de-excitation can proceed through the emission of monoenergetic axions carrying energies equal 
to the transition energy between the excited and ground states. These are 
magnetic dipole (M1) transitions,  discussed in Ref.~\cite{Donnelly:1978ty} for axion emission. List of nuclides with 
such low-lying M1 transitions in the keV energy range is given in Table~I of~\cite{Massarczyk:2021dje} along with the respective nuclear parameters and the transition energy. {Among these nuclides, $^{3}$He~\cite{Raffelt:1982dr,Borexino:2012guz,Lucente:2022esm}, $^{7}$Li~\cite{Derbin:2005xc,
Belli:2008zzb,Borexino:2008wiu,CAST:2009klq},  $^{57}$Fe~\cite{Moriyama:1995bz,
Krcmar:1998xn,CAST:2009jdc,CUORE:2012ymr}, and $^{83}$Kr~\cite{Jakovcic:2004sh,Gavrilyuk:2014mch} have significant abundances in the solar core.}


The sensitivity of helioscopes to various 
nuclear axion production channels has been recently assessed in 
Refs.~\cite{CAST:2009klq,CAST:2009jdc,DiLuzio:2021qct}. Data from the SNO 
experiment place a bound on the isotriplet axion--nucleon coupling $g_{3aN}$, 
excluding $g_{3aN} \gtrsim 2 \times 10^{-5}$ at 95\% confidence 
level~\cite{Bhusal:2020bvx}. The production of $5.49\,$MeV monochromatic solar axions from the M1 nuclear transition of excited $^{3}$He$^*$ produced via the nuclear 
fusion reaction $p + d \rightarrow {^{3}\rm{He}}^*$ was first calculated in 
Ref.~\cite{Raffelt:1982dr}. The Borexino experiment subsequently constrained 
such axions, placing bounds on the coupling combinations $(g_{3aN},\,g_{ae})$ 
and $(g_{3aN},\,g_{a\gamma})$~\cite{Borexino:2012guz}. More recently, the 
sensitivity of JUNO to these $5.49\,$MeV axions has been examined in ~\cite{Lucente:2022esm}, where it is shown that JUNO could improve upon the Borexino constraint by approximately an order of magnitude, owing to its superior energy resolution. 
Beyond solar axions, axions produced via M1 nuclear transitions of
$^{57}$Fe from the stellar population in nearby galaxies has recently been searched using X-ray data from the NuSTAR telescope~\cite{Ning:2025kyu}.

In this work, we search for solar axions emitted via the M1 nuclear 
de-excitation of $^{57}$Fe and $^{83}$Kr. Recent solar abundance studies~\cite{
2021A&A...653A.141A} have highlighted a significant abundance of $^{57}$Fe as 
well as a non-zero abundance of $^{83}$Kr in the solar core. 
%
{Both isotopes $^{57}$Fe and  $^{83}$Kr possess low-lying M1 excitation of $14.4\,$keV and $9.4\,$keV, respectively~\cite{Massarczyk:2021dje}.}
In the hot solar plasma, a fraction of 
these nuclei can be thermally excited to these low-lying states, which can subsequently 
de-excite via axion emission, yielding monochromatic axions {of the same energies.} 
%
Such solar axions have been searched for in several experiments, including CAST~\cite{CAST:2009jdc} and CUORE~\cite{CUORE:2012ymr} for the $14.4$ keV $^{57}$Fe channel via photon conversion and the axio-electric effect, respectively, and at the Baksan Neutrino Observatory for the $9.4$ keV $^{83}$Kr channel via resonant nuclear absorption~\cite{Gavrilyuk:2014mch}.

We investigate such axions using soft X-ray observations of the quiet Sun by the Chandrayaan-2 spacecraft. Chandrayaan-2 is a lunar orbiter of the Indian Space Research Organization 
(ISRO), launched in 2019 into an orbit around the Moon. Among its scientific 
payloads is the Solar X-ray Monitor (XSM), which records solar X-ray spectra 
in the $1$--$15\,$keV energy range. During the 2019--2020 solar 
minimum~\cite{Vadawale:2021soz,Vadawale:2021pis}, when solar activity was at 
its lowest, XSM recorded high-quality quiet-Sun soft X-ray spectra. The 
observed spectrum is dominated by two components: the quiescent coronal 
emission and the emission from flaring plasma. In addition, if axions are 
produced in the solar interior, they can convert into photons in the solar 
magnetic field, thereby contributing additional spectral features to the 
observed soft X-ray spectrum.
Previously  
Ref.~\cite{Kumar:2025yzl} searched for X-ray photons arising from the 
conversion of solar axions produced via the Primakoff effect, Compton 
scattering, and bremsstrahlung processes in the solar magnetic field, 
using Chandrayaan-2 data to set an upper bound on the axion-photon coupling. 
A similar analysis was previously performed using NuSTAR 
data~\cite{Ruz:2024gkl}. In the present work, we utilize the same 
Chandrayaan-2 solar soft X-ray dataset to search for monochromatic $14.4\,$keV and 
$9.4\,$keV photons arising from the conversion of $^{57}$Fe and $^{83}$Kr 
nuclear axions in the solar magnetic field, and in the absence of such spectral features, we derive upper limits on the 
axion-nucleon and axion-photon couplings.

The remainder of this paper is organized as follows. In Sec.~\ref{sec:solar_axion_production}, we discuss solar axion production and fluxes of the nuclei.  Sec.~\ref{sec:axion_photon_conversion} deals with conversion of axions to photons. Numerical analysis using the likelihood method for Chandrayaan-2 data and our limit on various axion couplings are discussed in Sec.~\ref{sec:limits}. Finally, our concluding remarks are presented in Sec.~\ref{sec:results}.

\section{Solar Axions from Nuclear Transition}
\label{sec:solar_axion_production}

The production and detection of axions are governed by their couplings to 
SM fields. The effective low-energy Lagrangian relevant for our 
analysis is given by~\cite{Lucente:2022esm}
\begin{align}
  \label{eq:AxionInter}
  \mathcal{L} = \frac{1}{2}\,\partial_{\mu} a\,\partial^{\mu} a - m^2_a a^2 
  - \frac{1}{4}\,\gagg\, a\, F_{\mu\nu}\widetilde{F}^{\mu\nu}
  - ia\,\bar{N}\gamma_5\left(g_{0aN} + \tau_3\, g_{3aN}\right)N\ ,
\end{align}
where the first two terms are the kinetic and mass terms of the axion field $a$, 
$F_{\mu\nu}$ and $\widetilde{F}^{\mu\nu}$ are the electromagnetic field 
strength tensor and its dual, and $N = (p,\, n)^{T}$ is the proton-neutron 
isospin doublet. The parameter $m_a$ denotes the axion mass, $g_{a\gamma\gamma}$ is 
the axion-photon coupling, and $g_{0aN}$ and $g_{3aN}$ are the iso-singlet and 
iso-triplet axion-nucleon couplings, respectively. As will be discussed later,  axion production in the solar core proceeds via the nucleon couplings $g_{0aN}$ 
and $g_{3aN}$, while detection relies on the conversion of axions into photons 
through the coupling $g_{a\gamma\gamma}$.

Apart from the standard de-excitation via photon emission, a nucleus in an excited state can alternatively decay to its ground 
state by emitting an axion. For the low-lying nuclear excited states of 
$^{57}$Fe and $^{83}$Kr we are concerned with, this process proceeds via a 
magnetic dipole (M1) transition, in which the axion carries away the full 
transition energy as a monochromatic signal. Specifically, for $^{57}$Fe the 
ground state has total angular momentum $J_0 = 1/2$ and the first excited state 
has $J_1 = 3/2$, while for $^{83}$Kr the ground state has $J_0 = 9/2$ and the 
first excited state has $J_1 = 7/2$~\cite{DiLuzio:2021qct}. In both cases, the change in angular 
momentum $|\Delta J| = 1$ is consistent with an M1 transition.

For such an M1 transition, the ratio of the transition rates for axion emission 
($\Gamma_a$) and photon emission ($\Gamma_\gamma$) is given 
by~\cite{Avignone:1988bv}
\begin{align}\label{eq:DecayRatio}
    \frac{\Gamma_a}{\Gamma_\gamma}=
\left( \frac{k_a}{k_\gamma} \right)^3
\frac{1}{2\pi\alpha_{\rm EM}}\frac{1}{1+\delta^2}
\left[ \frac{\beta \, g_{0aN} + g_{3aN}} 
{\left( \mu_0-\frac12 \right) \beta + \mu_3 -\eta} \right]^2\,.
\end{align}
Here, $k_a$ and $k_\gamma$ are the axion and photon momenta, respectively. 
For axion masses $m_a \ll 1\,$eV, we have $k_a = \sqrt{E_a^2 - m_a^2} 
\simeq 
E_a$ where $E_a$ is the axion energy and is equal to the energy of the M1 nuclear transition. Since the energy of the M1 nuclear transition is fixed, an axion and a photon emitted during the transition will have the same energy $ E_a = E_\gamma = k_\gamma$. Thus we have $k_a/k_\gamma \simeq 1$. The quantity 
$\alpha_{\rm EM}$ is the electromagnetic fine-structure constant, $\mu_0 = 
\mu_p + \mu_n \approx 0.88$ and $\mu_3 = \mu_p - \mu_n \approx 4.77$ are the 
isoscalar and isovector nuclear magnetic moments expressed in units of the 
nuclear magneton, and $\delta$ is the E2/M1 multipole mixing ratio of the 
transition. The parameters $\beta$ and $\eta$ encode the nuclear structure of 
the isotope and their values are tabulated in Ref.~\cite{Massarczyk:2021dje}. 
The branching ratio $\Gamma_a/\Gamma_\gamma$ (Eq.~\eqref{eq:DecayRatio}) quantifies the probability that a 
given nuclear transition proceeds via axion emission rather than photon emission.

For $^{57}$Fe, substituting the values $\delta = 0.002$, $\beta = -1.19$, and 
$\eta = 0.8$ from Ref.~\cite{Massarczyk:2021dje} into Eq.~\eqref{eq:DecayRatio}, 
we obtain
\begin{equation}
   \frac{\Gamma_a}{\Gamma_\gamma} = 1.82\,\left(-1.19\, g_{0aN} + g_{3aN}\right) 
   \equiv 1.82\, g_{aN,\,\mathrm{Fe}}^{\rm eff}\,,
\end{equation}
where we have defined the effective coupling $g_{aN,\,\mathrm{Fe}}^{\rm eff} = 
-1.19\, g_{0aN} + g_{3aN}$. 
{ It is important to note that $^{57}$Fe axion emissivity is strongly concentrated in the solar core, as the thermal population of the 14.4~keV excited state is exponentially sensitive to the local temperature,  as mentioned in  \cite{DiLuzio:2021qct}.
}

Similarly, for $^{83}$Kr, using $\delta = 0.002$, 
$\beta = -1$, and $\eta = 0.5$ from Ref.~\cite{Massarczyk:2021dje}, we find
\begin{equation}
   \frac{\Gamma_a}{\Gamma_\gamma} = -g_{0aN} + g_{3aN} 
   \equiv g_{aN,\,\mathrm{Kr}}^{\rm eff}\,,
\end{equation}
where $g_{aN,\,\mathrm{Kr}}^{\rm eff} = -g_{0aN} + g_{3aN}$. We note that 
the two effective couplings are numerically similar, 
$g_{aN,\,\mathrm{Fe}}^{\rm eff} \simeq g_{aN,\,\mathrm{Kr}}^{\rm eff}$, 
differing only through the isotope specific coefficient multiplying $g_{0aN}$.

\begin{figure}[t]
    \centering
    \includegraphics[width=0.7\linewidth]{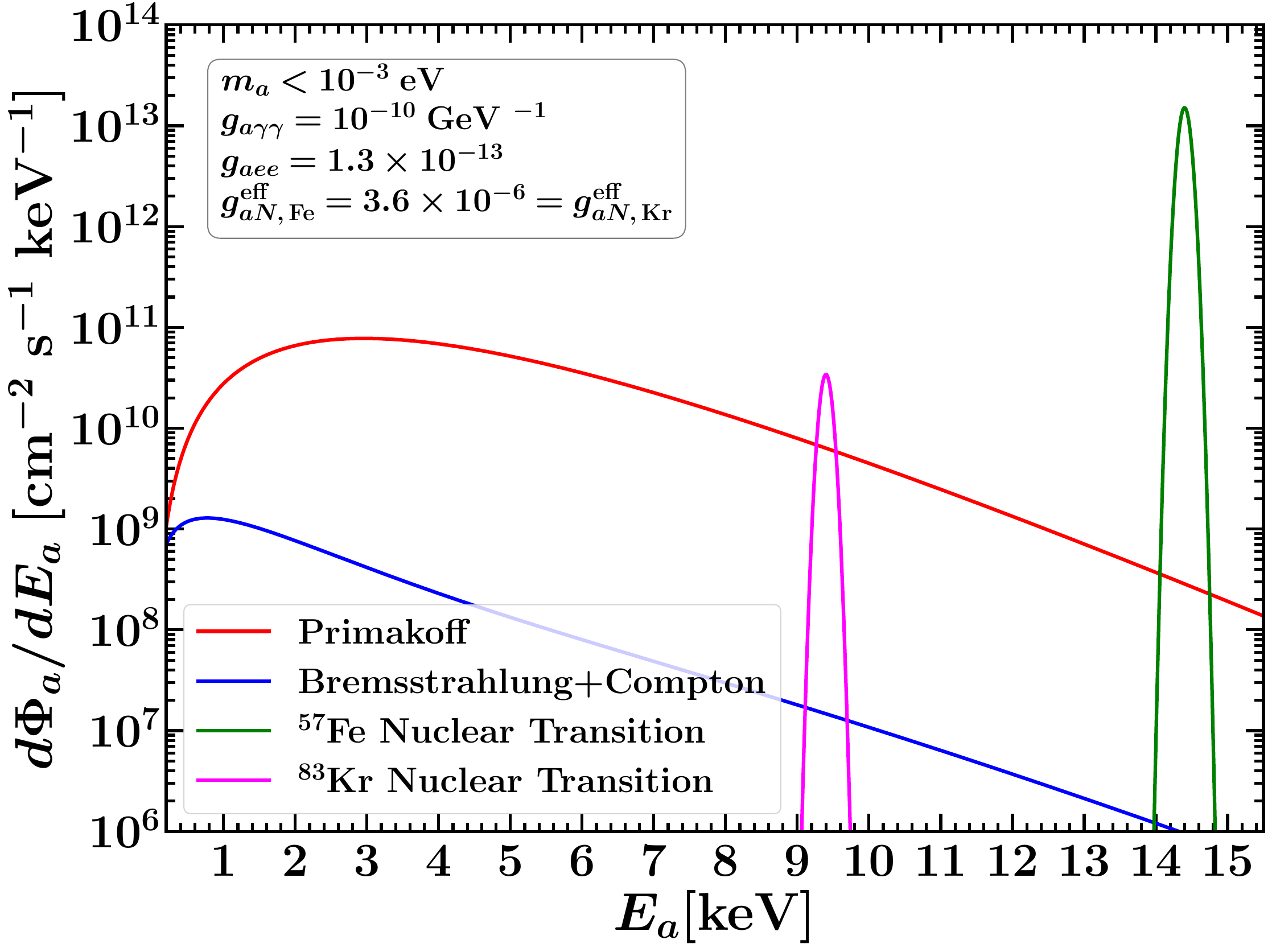}
    \caption{{
    Differential solar axion flux at Earth from various production mechanisms.
The continuous spectra include the Primakoff component (red) and the
electron-induced contributions -- bremsstrahlung and Compton (blue). The
line-like spectra arise from the 14.4~keV M1 nuclear transition of
$^{57}$Fe (green) and the 9.4~keV M1 nuclear transition of $^{83}$Kr
(magenta). For visualization purposes, the intrinsically monoenergetic
lines are shown broadened with a Gaussian of width corresponding to the
XSM energy resolution, illustrating their detector-broadened appearance.
    }}
    \label{fig:axion_flux}
\end{figure}

To calculate the flux of solar axions produced via the nuclear de-excitation of 
$^{57}$Fe and $^{83}$Kr, we follow the procedure outlined in 
Refs.~\cite{Moriyama:1995bz,DiLuzio:2021qct}. The axion emission rate per unit 
mass of the Sun is given by
\begin{equation}
    \mathcal{N}_a(r) = \mathcal{N}(r)\,\omega_1(T(r))\,\frac{1}{\tau_0}\,
    \frac{1}{1+\alpha}\,\frac{\Gamma_a}{\Gamma_\gamma}\,,
\end{equation}
where $\mathcal{N}(r)$ is the number of relevant nuclei per unit mass of the 
Sun, $\omega_1(T(r))$ is the thermal population fraction of the excited state at 
local temperature $T$, given by the Boltzmann factor
\begin{equation}
    \omega_1(T(r)) = \frac{(2J_1+1)\,e^{-E_1/T(r)}}{(2J_0+1) + (2J_1+1)\,e^{-E_1/T(r)}}\,,
\end{equation}
with $J_0$ and $J_1$ being the angular momenta of the ground and excited states, 
respectively, and $E_1$ the transition energy. The quantity $\tau_0$ is the 
mean lifetime of the nuclear excited state, $\alpha$ is the internal conversion 
coefficient which is the ratio of the number of nuclei undergoing de-excitation via photon emission to the number of nuclei undergoing de-excitation via electron emission, and $\Gamma_a/\Gamma_\gamma$ is the axion-to-photon branching 
ratio defined in Eq.~\eqref{eq:DecayRatio}. 

The number of relevant nuclei per unit mass of the Sun can be written as
\begin{equation}
    \mathcal{N}(r) = \frac{X_H(r)}{m_p}\,10^{\varepsilon-12}\,\mathcal{A}\,,
\end{equation}
where $X_H(r)$ is the local hydrogen mass fraction, $m_p$ is the proton mass,
$\varepsilon$ is the logarithmic photospheric abundance of the parent element
(Fe or Kr) relative to hydrogen in the standard astronomical scale
$\log_{10}(n_\mathrm{element}/n_H) + 12$ which we have adopted from~\cite{2021A&A...653A.141A}, and $\mathcal{A}$ is the isotopic abundance
fraction of the specific isotope ($^{57}$Fe or $^{83}$Kr) which we have adopted from~\cite{DiLuzio:2021qct}.

The axion emission rate per unit volume at radius $r$ is then
$\rho(r)\,\mathcal{N}_a(r)$, where $\rho(r)$ is the local mass density of
the solar plasma. The total axion luminosity of the Sun is obtained by
integrating this quantity over the full solar volume,
\begin{equation}
    L_a = \int_0^{R_\odot} \rho(r)\,\mathcal{N}_a(r)\,4\pi r^2\,\mathrm{d}r\,.
\end{equation}
Assuming isotropic emission, the differential flux of solar axions at Earth
(distance $d_\odot$ from the Sun) is
\begin{equation}
    \Phi_a = \frac{L_a}{4\pi\,d_\odot^2}\,.
\end{equation}
Combining the expressions above, the total solar axion flux at Earth from the
nuclear de-excitation of a given isotope is
\begin{align}\label{eq:flux}
    \Phi_a
    = \frac{a\,10^{\varepsilon-12}}{4\pi\,d_\odot^2}
    \cdot\frac{1}{\tau_0\,(1+\alpha)}
    \cdot\frac{\Gamma_a}{\Gamma_\gamma}
    \cdot 
    \underbrace{%
      \int_0^{R_\odot}
      \omega_1(T(r))\,
      \frac{\rho(r)\,X_H(r)}{m_p}\,
      4\pi r^2\,\mathrm{d}r
    }_{\displaystyle\equiv\,\mathcal{I}}\,,
\end{align}
where all nuclear parameters are isotope-specific and the dimensionless
integral $\mathcal{I}$ encodes the thermal and compositional structure of the
solar interior. $\mathcal{I}$ is evaluated numerically using the B23-AGSS09
standard solar model from~\cite{Herrera_Serenelli_2023}, which provides radial
profiles of $T(r)$, $\rho(r)$, and $X_H(r)$.

Substituting the respective nuclear parameters into Eq.~\eqref{eq:flux},
the axion fluxes from the $14.4\,\mathrm{keV}$ transition of $^{57}$Fe and
the $9.4\,\mathrm{keV}$ transition of $^{83}$Kr at Earth read as
\begin{align}
    \Phi_a^{\rm Fe}
    &= \Phi_0^{\rm Fe}
       \left(g_{aN,\,\mathrm{Fe}}^{\rm eff}\right)^2\,,
    \label{eq:fluxFe}\\[6pt]
    \Phi_a^{\rm Kr}
    &= \Phi_0^{\rm Kr}
       \left(g_{aN,\,\mathrm{Kr}}^{\rm eff}\right)^2\,,
    \label{eq:fluxKr}
\end{align}
where 
\begin{align}
    \Phi_0^{\rm Fe} &= 2.18 \times 10^{23}\,\,\text{cm}^{-2}\,\text{s}^{-1}\,\, {\rm and}\,\,  \Phi_0^{\rm Kr} = 4.92 \times 10^{20}\,\,\text{cm}^{-2}\,\text{s}^{-1}.
    \label{eq:normalization_flux}
\end{align}

As is evident from Eq.~\eqref{eq:normalization_flux}, the $^{83}$Kr axion flux is
roughly three orders of magnitude smaller than that of $^{57}$Fe. This
suppression is predominantly due to the much lower photospheric abundance of
krypton relative to iron, $\varepsilon_{\rm Kr} \approx 3.12$ versus
$\varepsilon_{\rm Fe} \approx 7.46$~\cite{2021A&A...653A.141A},
which translates directly into a correspondingly smaller number of
emitting nuclei in the solar interior.

These monoenergetic axion fluxes are displayed in Fig.~\ref{fig:axion_flux},
where the green and magenta lines correspond to the $14.4\,\mathrm{keV}$
and $9.4\,\mathrm{keV}$ lines from the M1 nuclear transitions of
$^{57}\mathrm{Fe}$ and $^{83}\mathrm{Kr}$, respectively. For comparison,
we also show the continuous axion spectra from the Primakoff process
(red) and the combined electron bremsstrahlung and Compton contributions
(blue).


\section{X-ray flux from Axion-Photon Conversion Outside the Sun}
\label{sec:axion_photon_conversion}
A fraction of the axions produced in the solar interior can convert into
photons as they traverse the Sun's external magnetic field, via the
inverse Primakoff process~\cite{Raffelt:1987im}. For the monoenergetic
axions considered here, this conversion yields a flux of X-ray photons
at the same energies as the parent nuclear transitions, $14.4\,\mathrm{keV}$
for $^{57}$Fe and $9.4\,\mathrm{keV}$ for $^{83}$Kr, arriving at Earth
as sharp spectral lines superimposed on the solar X-ray background.
An X-ray telescope observing the Sun can therefore search for these
line features in the detected spectrum, and the non-observation of
any such excess can be used to place constraints on the product of the
axion-nucleon and axion-photon couplings, $|g_{aN}^{\rm eff}\times \gagg|$.

To calculate the intensity of these X-rays, we start by calculating the conversion probability of axions into photons while traveling through a magnetized plasma. In the presence of an external magnetic field $\mathbf{B}(\ell)$, in this case the solar magnetic field, axions mix with the component of the photon field which is polarized parallel to the external magnetic field~\cite{Raffelt:1987im} resulting in axion-photon oscillation. The equation of motion for  axions in a magnetized plasma is given by~\cite{Raffelt:1987im}
\begin{equation}
	\bigg( (E_a -i\partial_\ell)\mathbb{I} + \hat M(\ell)\bigg)
	\begin{pmatrix}
		A_\parallel(\ell)\\
		a(\ell)\\
	\end{pmatrix} = 0\,,
	\label{eq:EqofMotion}
\end{equation}
where 
$\ell$ is the distance traveled by the photon/axion in the magnetized plasma,
$A_{\parallel}$ represents the photon polarization state parallel  to $\mathbf{B}_{\rm T}(\ell)$ where, $\mathbf{B}_{\rm T}(\ell)$ is the transverse projection of the external magnetic field $\mathbf{B}(\ell)$, and
$\hat M(\ell)$ represents a $2 \times 2$ mixing matrix, given by
\begin{equation}
	\hat M(\ell) = 
	\begin{pmatrix}
		-\frac{\omega_p^2(\ell)}{2\omega} - i \dfrac{\Gamma(\ell)}{2}         &g_{a\gamma\gamma}\frac{B_T(\ell)}{2}\\
		g_{a\gamma\gamma}\frac{B_T(\ell)}{2} &-\frac{m_a^2}{2\omega}\\
	\end{pmatrix}\,,
\end{equation}
 where $\omega_p(\ell) = \sqrt{4\pi\, \alpha \,n_e(\ell)/m_e}$ represents the plasma frequency of the photon with $n_e(\ell)$ being the electron density of the plasma. In this case the plasma present in the solar atmosphere and $B_{T}(\ell) = |\mathbf{B}_{\rm T}(\ell)|$. Here, $\Gamma(\ell) = \sum_{Z=1}^{30} n_Z(\ell)\, \sigma_Z$ is the
 absorption coefficient of the X-ray photon in the plasma with $n_Z$ being the number density of element of atomic number $Z$ and $\sigma_Z$ being the total attenuation cross-section of a photon passing through a gas of atomic number $Z$. 
To calculate the conversion probability, we solve the equation of
motion~\eqref{eq:EqofMotion} subject to initial conditions at $\ell = 0$ which in our case we consider to be the solar surface.
The solution at some location at a distance $\ell = h$ from the initial point is given by the path-ordered
transfer matrix~\cite{Raffelt:1987im,Manzari:2024jns},
\begin{align}
\label{eq:homsoln}
\begin{pmatrix}
    A_\parallel(h)\\
    a(h)
\end{pmatrix}
= {\cal P}_\ell\Bigg[
    \exp\!\left(
        -i E_a h\,\mathbb{I}
        - i \int_0^h \hat{M}(\ell)\, \mathrm{d}\ell
    \right)
\Bigg]
\begin{pmatrix}
    A_\parallel(0)\\
    a(0)
\end{pmatrix},
\end{align}
where $A_\parallel(0)$ and $a(0)$ are the initial photon and axion
field amplitudes, respectively, $\mathbb{I}$ is the $2\times 2$
identity matrix, $\hat{M}(\ell)$ is the mixing matrix encoding the
local plasma and magnetic field properties, and $\mathcal{P}_\ell$
denotes path ordering along the propagation direction.

Taking the initial conditions at the solar surface to be $A_\parallel(0) = 0$
and $a(0) = 1$, i.e.\ a pure axion state with no photon admixture, the
transfer matrix in Eq.~\eqref{eq:homsoln} can be evaluated perturbatively
in powers of $\gagg$~\cite{Raffelt:1987im}. To first order in
perturbation theory, the axion-to-photon conversion probability reads

\begin{equation}
\label{eq:osc_prob}
    P_{a\gamma}(h,\,E_a,\,m_a,\,\gagg)
    = \frac{\gagg^2}{4}\,
    e^{-\int_0^h \mathrm{d}\ell\;\Gamma(\ell)}
    \left|
        \int_0^h \mathrm{d}\ell\;B_T(\ell)\,
        \exp\!\left[
            i\int_0^\ell \mathrm{d}\ell'\,q(\ell')
            + \frac{1}{2}\int_0^\ell \mathrm{d}\ell'\,\Gamma(\ell')
        \right]
    \right|^2,
\end{equation}

where $q(\ell) = \bigl(\omega_p^2(\ell) - m_a^2\bigr)/2E_a$ is the momentum
transfer between the axion and photon.

\begin{figure}[t]
    \centering
   \includegraphics[width=0.7\linewidth]{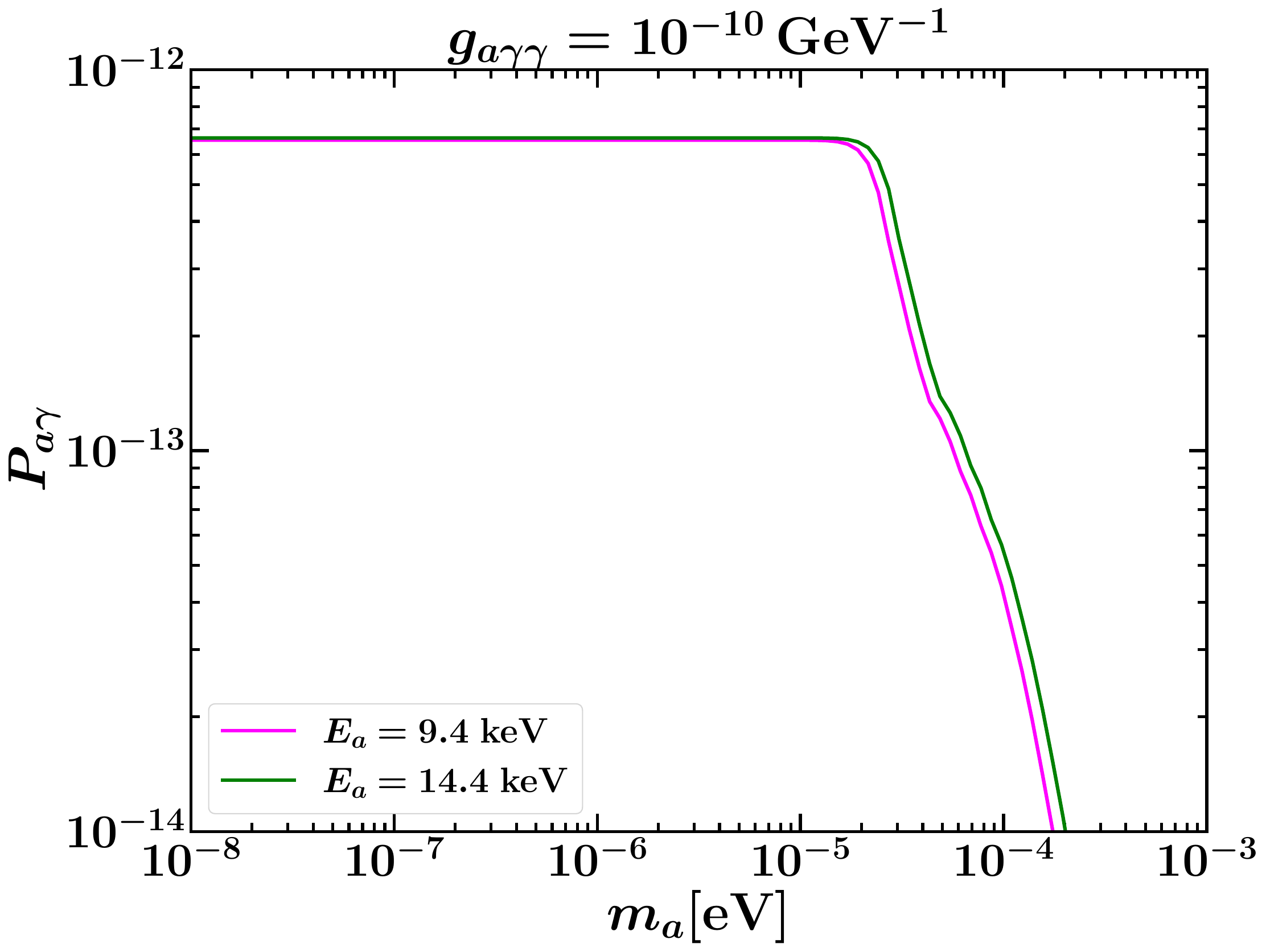}
    \caption{ Axion to photon oscillation probability at a far away distance from the Sun as a function of $m_a$  for $E_a = 9.4$ keV (magenta solid line) and $E_a = 14.4$ keV (green solid line.) 
    }
    \label{fig:osc_prob}
\end{figure}

To obtain the X-ray signal intensity at Earth, we numerically evaluate
Eq.~\eqref{eq:osc_prob} using the {electron density profile and the median solar magnetic field strength profile\footnote{See appendix for discussion}} adopted in Ref.~\cite{Ruz:2024gkl}, which extend to a heliocentric
distance of $h = 29\,R_\odot$ and we therefore perform the integration out to this distance.
The photon absorption coefficient $\Gamma(\ell)$ is computed by summing
contributions from all elements up to atomic number $Z = 30$, weighted
by their local abundances in the solar atmosphere as tabulated in the
\texttt{CHIANTI} atomic database~\cite{2024ApJ...974...71D,2021ApJ...909...38D},
accessed via the \texttt{ChiantiPy} interface~\cite{ChiantiPy:2022xx}.
For each element, we use the total photon attenuation cross-section
including coherent scattering from the NIST \texttt{XCOM}
database~\cite{xcom}.

{Fig.~\ref{fig:osc_prob} shows the conversion probability of axion to photon $P_{a\gamma}$,
evaluated at $h = 29\,R_\odot$, as a function of the axion mass $m_a$
for the two transition energies, namely $14.4\,\mathrm{keV}$ (green) and $9.4\,\mathrm{keV}$ (magenta), respectively. In the low-mass regime,
$m_a \lesssim 10^{-5}\,\mathrm{eV}$, the conversion probability is
constant,
while for $m_a \gtrsim 10^{-5}\,\mathrm{eV}$ it falls rapidly.}
This behavior of the oscillation probability critically depends on the momentum exchange between the axion and the photon $q(\ell)$, which dictates the phase difference between the axion and the photon waves. 
The momentum transfer defines the oscillation length, $L_{\text{osc}} = 2\pi / q$, over which the waves drift out of phase. Optimal conversion requires that the spatial extent of the magnetic field is much smaller than the oscillation length, ensuring constructive interference (coherence).

For $m_a \lesssim 10^{-5}\,\mathrm{eV}$
the relevant conversion region lies in the solar corona ($h \gtrsim 10^{-2}\,R_\odot$,
see Fig.~2 of Ref.~\cite{Ruz:2024gkl}), where the free-electron density is very
low. As a consequence, the plasma frequency is extremely small,
$\omega_p \sim 10^{-8}\,\mathrm{eV}$, as well as the photon absorption coefficient
$\Gamma$ is negligible, rendering the medium essentially transparent to X-rays.
Since both $m_a$ and $\omega_p$ are far below the photon energy, the momentum
transfer $q \to 0$ and the oscillation length $L_{\rm osc} = 2\pi/q$ become
much larger than the physical extent of the corona. The phase factor in
Eq.~\eqref{eq:osc_prob} therefore approaches unity, meaning the axion and photon
waves remain perfectly in phase across the entire coronal volume. In this fully
coherent, low-absorption regime, the conversion probability simplifies to
\begin{equation}
    P_{a\gamma} \;\propto\; \left( \gagg \int_0^h \mathrm{d}\ell\; B_T(\ell) \right)^2,
\end{equation}
which is independent of $m_a$, giving rise to the characteristic flat plateau
seen in Fig.~\ref{fig:osc_prob}. Although the coronal field is weak, the large integration length and perfect
coherence are sufficient to dominate the signal.

For $m_a \gtrsim 10^{-5}\,\mathrm{eV}$, the conversion probability is suppressed
by two competing effects. Firstly for these masses the resonance condition $m_a \simeq \omega_p$
is satisfied in the chromosphere or photosphere (see Fig.~2 of
Ref.~\cite{Ruz:2024gkl}), where the magnetic field is strong
($B \sim 10^2\,\mathrm{G}$) but the plasma density and thereby the free electron and ion density is also very high. Since the
absorption coefficient scales with the free electron and ion density the optical depth in these layers is very large, and any photons produced by resonant
conversion are rapidly reabsorbed before escaping. Secondly, in the overlying
transparent corona where $m_a \gg \omega_p$, the momentum transfer grows as
$q \approx m_a^2/2E_a$, driving $L_{\rm osc}$ well below the coronal scale.
The axion and photon waves quickly dephase, and the conversion becomes incoherent.
These two effects namely opacity suppression at the deep-atmosphere resonance and phase
mismatch in the corona, combine to produce the steep fall-off of $P_{a\gamma}$
at higher masses.
The threshold at $m_a \sim 10^{-5}\,\mathrm{eV}$ therefore marks the physical
boundary between the chromosphere and the corona, and the shape of
$P_{a\gamma}(m_a)$ is a direct map of the solar atmosphere's structure.

Given the conversion probability, the X-ray photon flux at Earth from
axion-photon conversion is obtained by multiplying the solar axion flux
by $\hat{P}_{a\gamma}$,
\begin{equation}
    \Phi_\gamma(E) = \Phi_a \times \delta(E - E_a)
                   \times \hat{P}_{a\gamma}(E,\,m_a,\,\gagg)\,,
    \label{eq:photon_flux}
\end{equation}
where $E_a$ is the energy of the monoenergetic axions and
\begin{equation}
    \hat{P}_{a\gamma}(E,\,m_a,\,\gagg) \equiv P(h = 29\,R_\odot,\,E,\,m_a,\,\gagg)
\end{equation}
is the conversion probability evaluated at the outer boundary of the
integration domain. The $\delta$-function encodes the line-like shape of the spectra.

{
\section{XSM on-board Chandrayaan-2: Data set and background}\label{sec:xsm_details}
Chandrayaan-2 is an ISRO-developed lunar exploration mission launched in July 2019 to study the composition of the lunar surface. During the 2019-2020 solar minimum, the Solar X-ray Monitor (XSM) onboard the Chandrayaan-2 measured the integrated solar X-ray spectrum originating from the entire solar disk in the $1-15$ keV energy range~\cite{2020CSci..118...45S,2020SoPh..295..139M}.
Similar to~\cite{Kumar:2025yzl}, in this work we use the total X-ray spectrum reported in~\cite{Vadawale:2021soz}, which has been obtained from the XSM observations of the quiet Sun
during its first two observing seasons. For the first season, observations were made during September 2019 to November 2019 and the second  during February 2020 to May 2020, when the solar activity was at its minimum and  no active regions or micro-flares were present.

This X-ray spectrum consists of two components: X-rays from the quiet Sun and a background component.
The quiet Sun spectrum is dominated
by the quiescent coronal emission in the low energy range up to $\sim$3 keV~\cite{Vadawale:2021soz}, while at higher energies the coronal emission is negligibly small. Thus, to get rid of the quiet Sun X-rays, in our analysis we used the XSM recorded spectrum above 3.4 keV to search for X-ray produced from axion conversion.
The background component, on the other hand, primarily consists of Cosmic X-ray Background (CXB),
charged particle background, and emission from other bright astrophysical sources.
Although the CXB component is stable, the measured CXB by XSM varies. Some part of the XSM field-of-view (FOV) is usually obstructed by the Moon blocking CXB emission, and the blocked fraction varies due to the variation in the orbital altitude of the spacecraft.
The charged particle background has temporal variations due to passage through geotail regions, solar energetic particle (SEP) events, and Coronal Mass Ejection (CME) events, which are also recorded by the upper level discriminator (ULD) counts of the XSM instrument~\cite{2020SoPh..295..139M}.
As the spacecraft attitude varies, bright astrophysical sources, Sco X-1 being the brightest source in this energy range,
enter within the XSM FOV for some periods and contribute to the background.
Such periods of varying background with high ULD counts and presence of the Sco X-1 within the FOV were ignored
while generating the quiet Sun X-ray spectrum in~\cite{Kumar:2025yzl}, to ensure that the background was as clean as possible, resulting in the net effective exposure of $3.72 \times \mathrm{10^6}$ s.

To obtain our limits, we considered the three different background subtraction schemes used in~\cite{Kumar:2025yzl} to obtain the residual X-ray emission from the Sun:
\begin{itemize}
\item {\bf Case 1 (Conservative):} Only the CXB component~\cite{2010A&A...512A..49T}, taking into account the fraction of FOV obstructed by the Moon, is subtracted from the observed XSM spectrum. The resulting residuals give the most conservative limit.

\item {\bf Case 2 (Realistic):} The background observed when the Sun is out of the FOV is subtracted from the quiet Sun observations, after excluding periods when any of the 13 bright sources identified from MAXI all-sky monitor data were within the XSM FOV, yielding an effective quiet Sun exposure of
$2.15 \times 10^6$ s. Since this procedure also removes the charged particle background and other contaminating sources, it gives a more realistic estimate of the residual solar x-ray spectrum.

\item {\bf Case 3 (Optimistic):} The entire observed spectrum from Case 1 is assumed to arise from the background. An empirical model is fit to the observed spectrum, and the best-fit model is subtracted from the data to obtain the residual, corresponding to the case where the background is perfectly modeled.
\end{itemize}

In Fig.~\ref{fig:signal_vs_residual}, as a case study we show only Case 3 (see the figure caption for details).
%
%
The black dots are the residuals after background subtraction, while the error bars denote $1\sigma$ uncertainties arising from Poisson errors in the observed counts. However, all three residual spectra are used in Sec.~\ref{sec:limits} to derive limits on the axion-photon coupling, allowing us to bracket the impact of background modeling systematics on our final result.

}

\section{Limits from Chandrayaan-2}\label{sec:limits}

\begin{figure}[t!]
    \centering
      \includegraphics[width=0.48\linewidth]{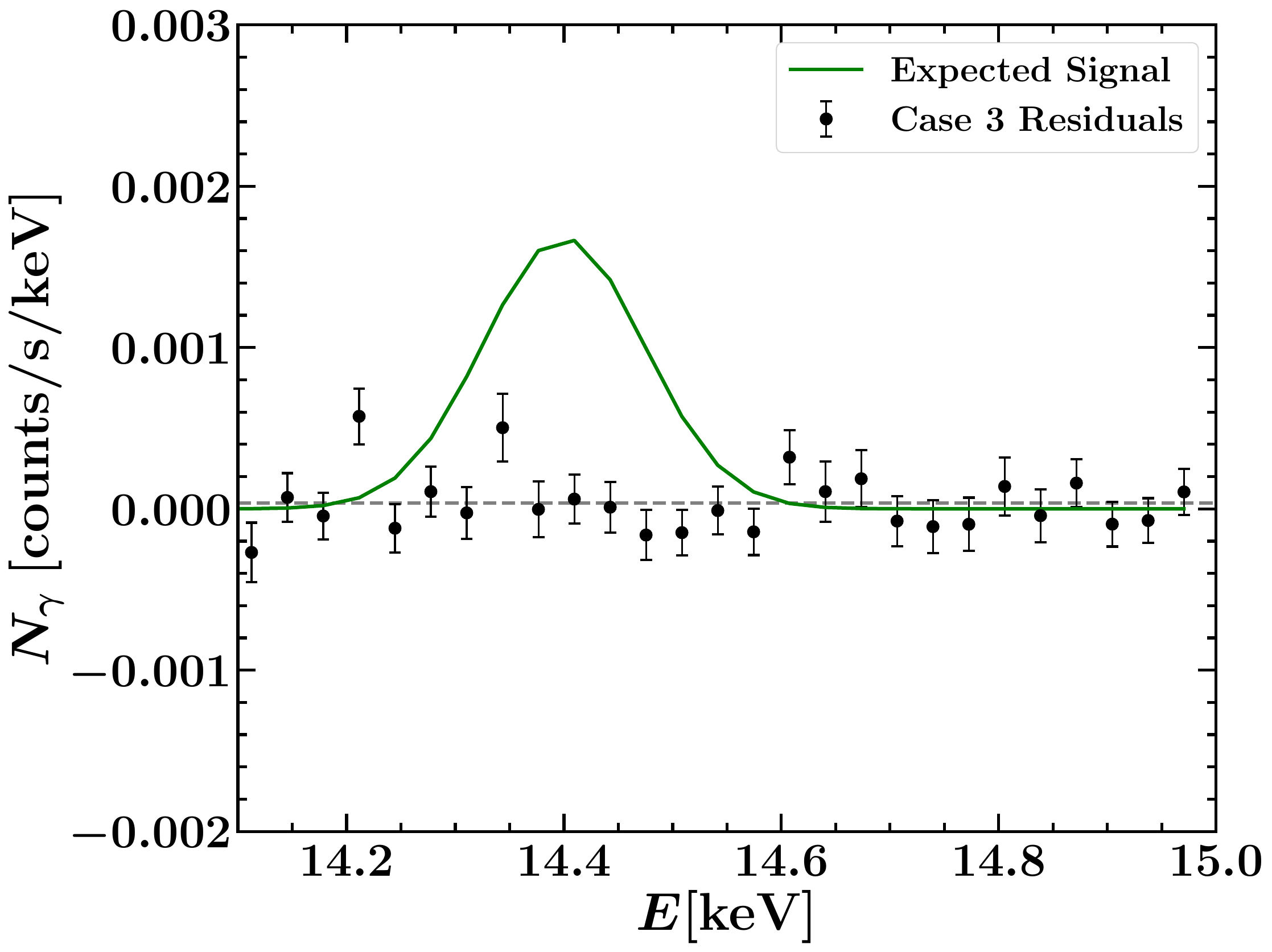}
    \includegraphics[width=0.48\linewidth]{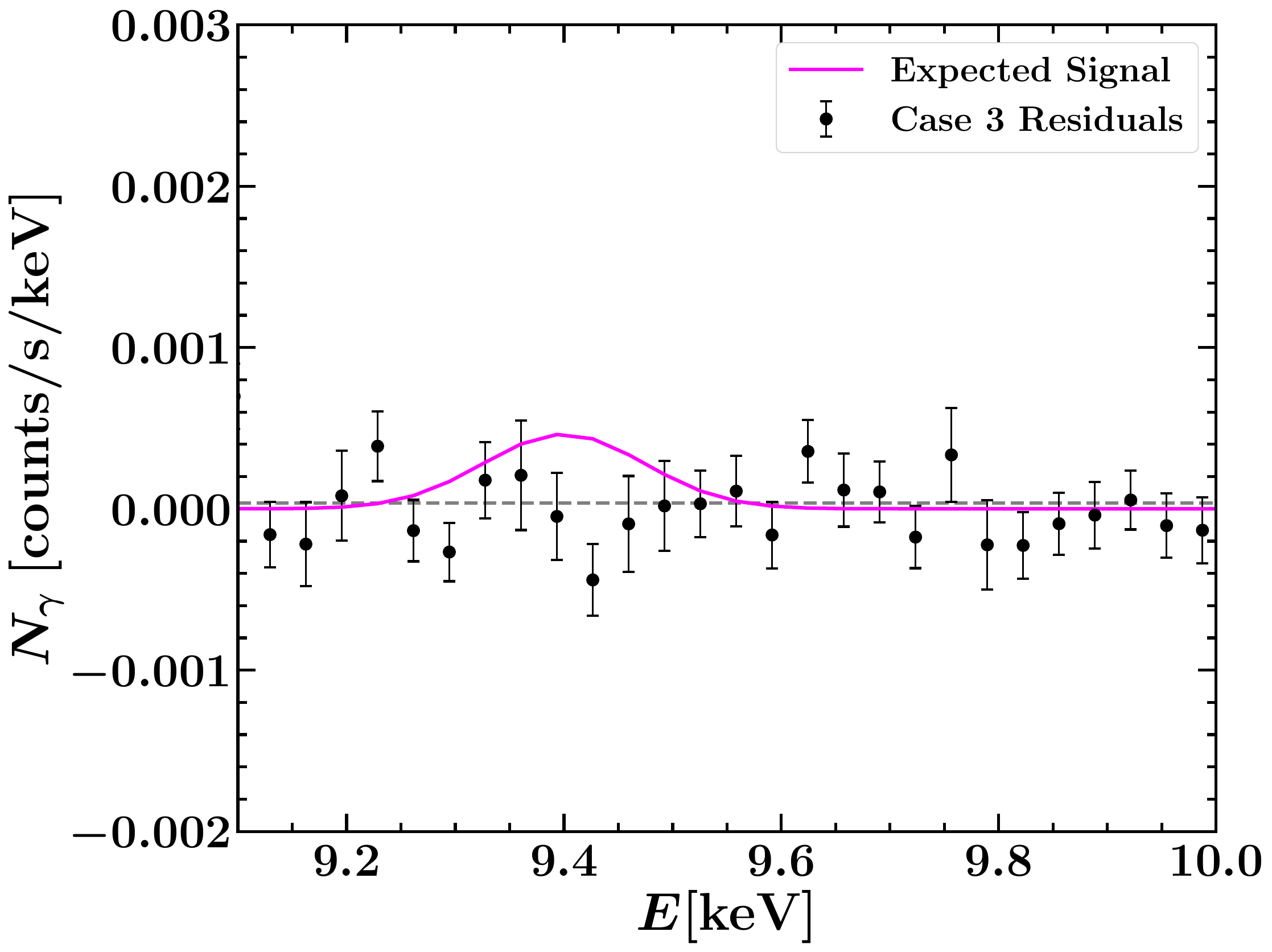}
    \caption{\footnotesize The expected solar axion-induced photon signal from the 14.4~keV
$^{57}$Fe transition for $|g_{aN,\,\rm Fe}^{\rm eff}\times\gagg| =
10^{-16}~\mathrm{GeV}^{-1}$ (\textit{\textbf{left panel}}) and from the
9.4~keV $^{83}$Kr transition for $|g_{aN,\,\rm Kr}^{\rm eff}\times\gagg| =
10^{-15}~\mathrm{GeV}^{-1}$ (\textit{\textbf{right panel}}), shown together
with the residual spectrum after background subtraction.  Case~3.   has been considered as representative of the three background-subtraction schemes (see Sec.~\ref{sec:xsm_details}). The axion signal is shown after convolution with
the detector energy resolution, while the residuals represent the
difference between the observed spectrum and the best-fit background model
in the corresponding energy window.
}
    \label{fig:signal_vs_residual}
\end{figure}

To obtain the limits on the axion couplings, we adopt the following procedure. First, we construct the theoretical signal, by taking the X-ray photon flux
$\Phi_\gamma(E)$ derived in Eq.~\eqref{eq:photon_flux}, which contains a 
$\delta$-function encoding the monoenergetic nature of the emission. To 
account for the finite energy resolution of the XSM, we replace this 
$\delta$-function with a Gaussian of width 
$\sigma = \mathrm{FWHM}/(2\sqrt{2\ln 2})$, with 
$\mathrm{FWHM} = 175~\mathrm{eV}$~\cite{2020CSci..118...45S}. 
The expected photon counts in the $i$-th energy bin are then obtained 
by folding with the energy-dependent effective area $A_{\rm eff}(E)$, 
\begin{equation}
    N_{\gamma,\,\rm exp}(E_i) = 
    \Phi_a \cdot \hat{P}_{a\gamma}(E_i,\,m_a,\,\gagg)
    \cdot \frac{A_{\rm eff}(E_i)}{\sqrt{2\pi}\,\sigma}
    \exp\!\left[-\frac{(E_i-E_a)^2}{2\sigma^2}\right],
    \label{eq:N_exp}
\end{equation}
where $E_a = 14.4\,(9.4)~\mathrm{keV}$ for $^{57}$Fe ($^{83}$Kr) and $\Phi_a$ is given in Eq.~\eqref{eq:normalization_flux}. { Note that the effective area $A_{\rm eff}(E_i)$ is given by the product of the detector efficiency and geometric area of
XSM.}
Denoting the coupling-dependent overall normalization as
\begin{align}
    N_S(E_i) &\;\equiv\; 
    \Phi_a \cdot \hat{P}_{a\gamma}(E_i,\,m_a,\,\gagg) \cdot A_{\rm eff}(E_i) \nn \\
    &= (g_{aN,\, \rm Fe/Kr}^{\rm eff} \times \gagg )^2 \cdot \Phi_0^{\rm Fe/Kr} \cdot \hat{P}_{a\gamma}(E_i,\,m_a,\,\gagg=1) \cdot A_{\rm eff}(E_i),
    \label{eq:NS_def}
\end{align}
we have
\begin{equation}
    N_{\gamma,\,\rm exp}(E_i) = N_S(E_i) \cdot 
\frac{1}{\sqrt{2\pi}\,\sigma}
\exp\left[\dfrac{-(E_i-E_a)^2}{2\sigma^2}\right].
\end{equation}
The expected signal $N_{\gamma,\,\rm exp}(E_i)$ obtained from
Eq.~\eqref{eq:N_exp} is shown in Fig.~\ref{fig:signal_vs_residual} for the
$^{57}$Fe (left panel) and $^{83}$Kr (right panel) transitions, for
representative coupling values of
$|g_{aN,\,\rm Fe}^{\rm eff}\times\gagg| = 10^{-16}~\mathrm{GeV}^{-1}$ and
$|g_{aN,\,\rm Kr}^{\rm eff}\times\gagg| = 10^{-15}~\mathrm{GeV}^{-1}$,
respectively, alongside the Case~3 residual spectrum in the corresponding
energy window.

Now given the residual count rate per unit energy $N_{\gamma,\,\rm obs}(E_i)$
for a particular background subtraction scheme, we construct a Gaussian 
likelihood over all energy bins in the range $3.4$--$15~\mathrm{keV}$,
\begin{equation}
    \mathcal{L}(|g_{aN}^{\rm eff}\times \gagg|;\,m_a) \propto 
    \prod_i \exp\!\left[-\frac{1}{2}
    \left(\frac{N_{\gamma,\,\rm exp}(E_i) - 
    N_{\gamma,\,\rm obs}(E_i)}{\tilde{\sigma}(E_i)}\right)^2\right],
    \label{eq:likelihood}
\end{equation}
where $\tilde{\sigma}(E_i)$ denotes the uncertainty in the observed counts.
{We adopt a Gaussian likelihood, Eq.~\eqref{eq:likelihood} since the fit is
performed on the background subtracted residual count rate
$N_{\gamma,\,\rm obs}(E_i)$ rather than on raw photon counts. {Additionally we treat the uncertainties $\tilde{\sigma}(E_i)$ in Eq.~\eqref{eq:likelihood}
as uncorrelated across energy bins, and do not include a separate
energy-calibration uncertainty, as none is available for the XSM dataset
used in this analysis}.
}

\begin{figure}[t!]
    \centering
    \includegraphics[width=0.48\linewidth]{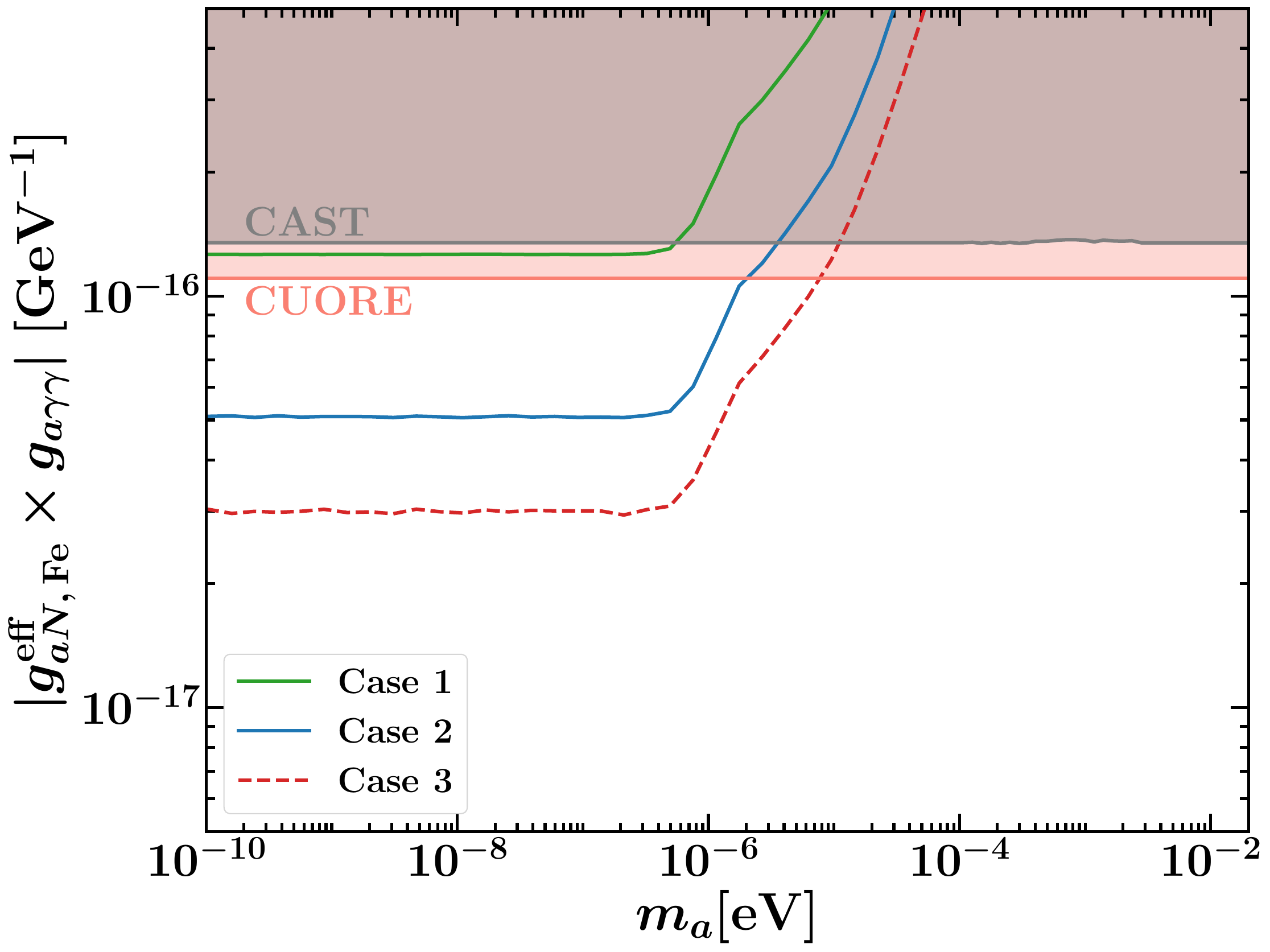}
    \includegraphics[width=0.48\linewidth]{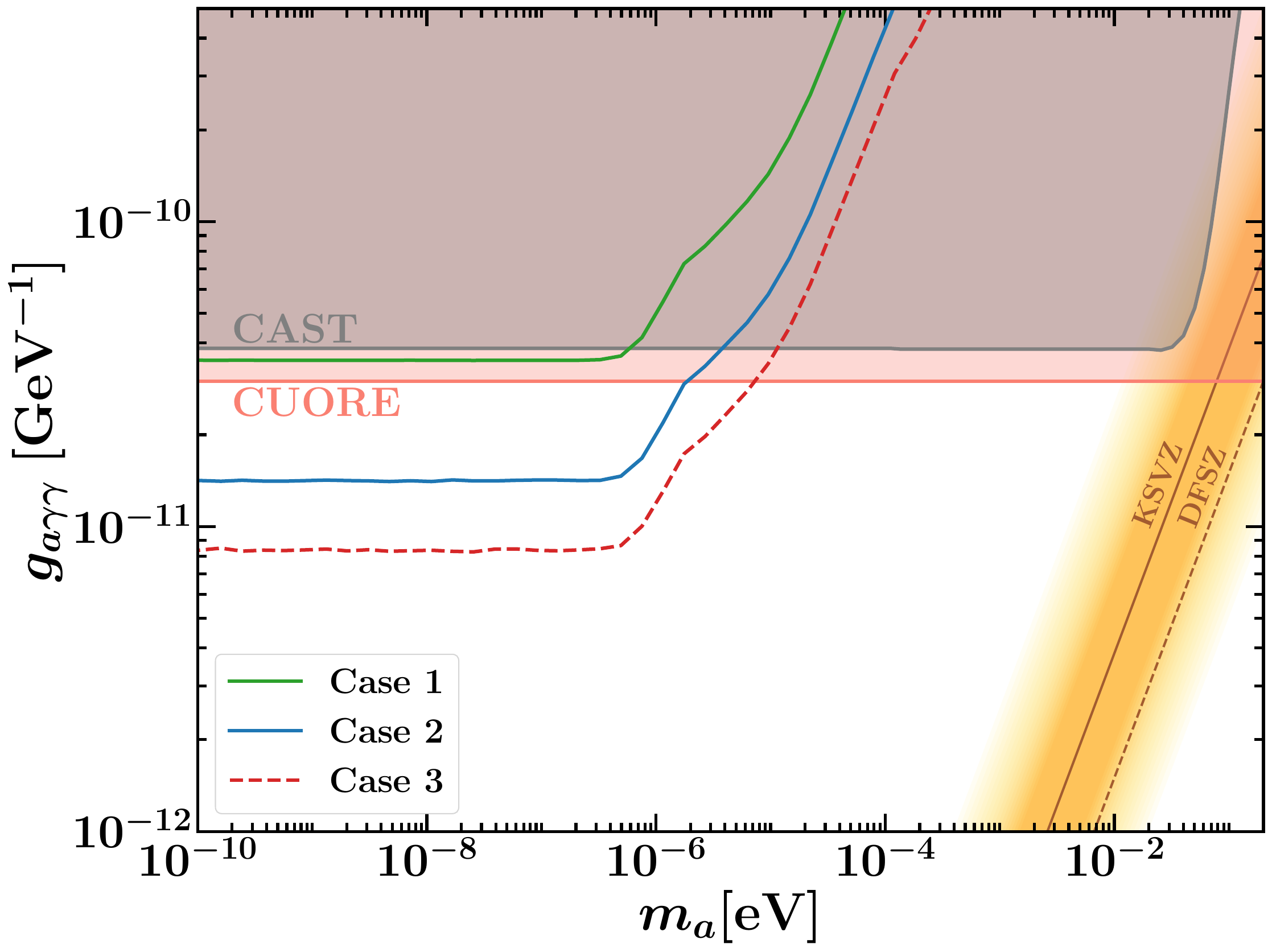}
    \caption{\footnotesize Upper limits at 95\% confidence level on the product $|g_{aN,\,\mathrm{Fe}}^{\rm eff} \times \gagg|$ (left panel) and  $\gagg$ for $g_{aN,\,\mathrm{Fe}}^{\rm eff} = 3.6 \times 10^{-6}$(right panel) as a function of the axion mass, derived from Chandrayaan-2 XSM observations. The constraint is obtained from the non-observation of an excess at the 14.4 keV $^{57}$Fe nuclear transition in the solar X-ray spectrum. The yellow region represents the QCD axion band~\cite{AxionLimits}.
    }
    \label{fig:constraint_Fe}
\end{figure}

\begin{figure}[t!]
    \centering
    \includegraphics[width=0.48\linewidth]{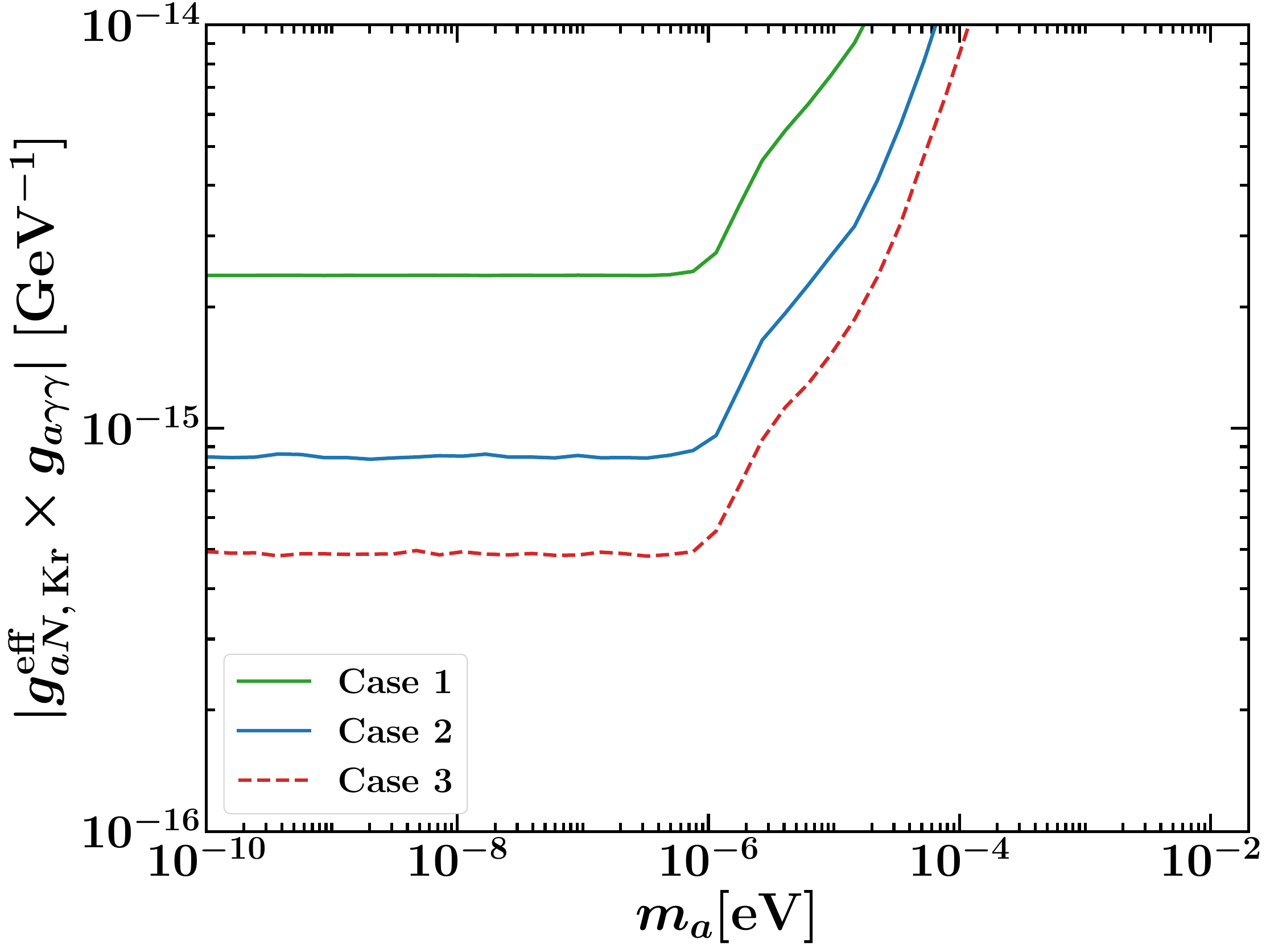}
    \includegraphics[width=0.48\linewidth]{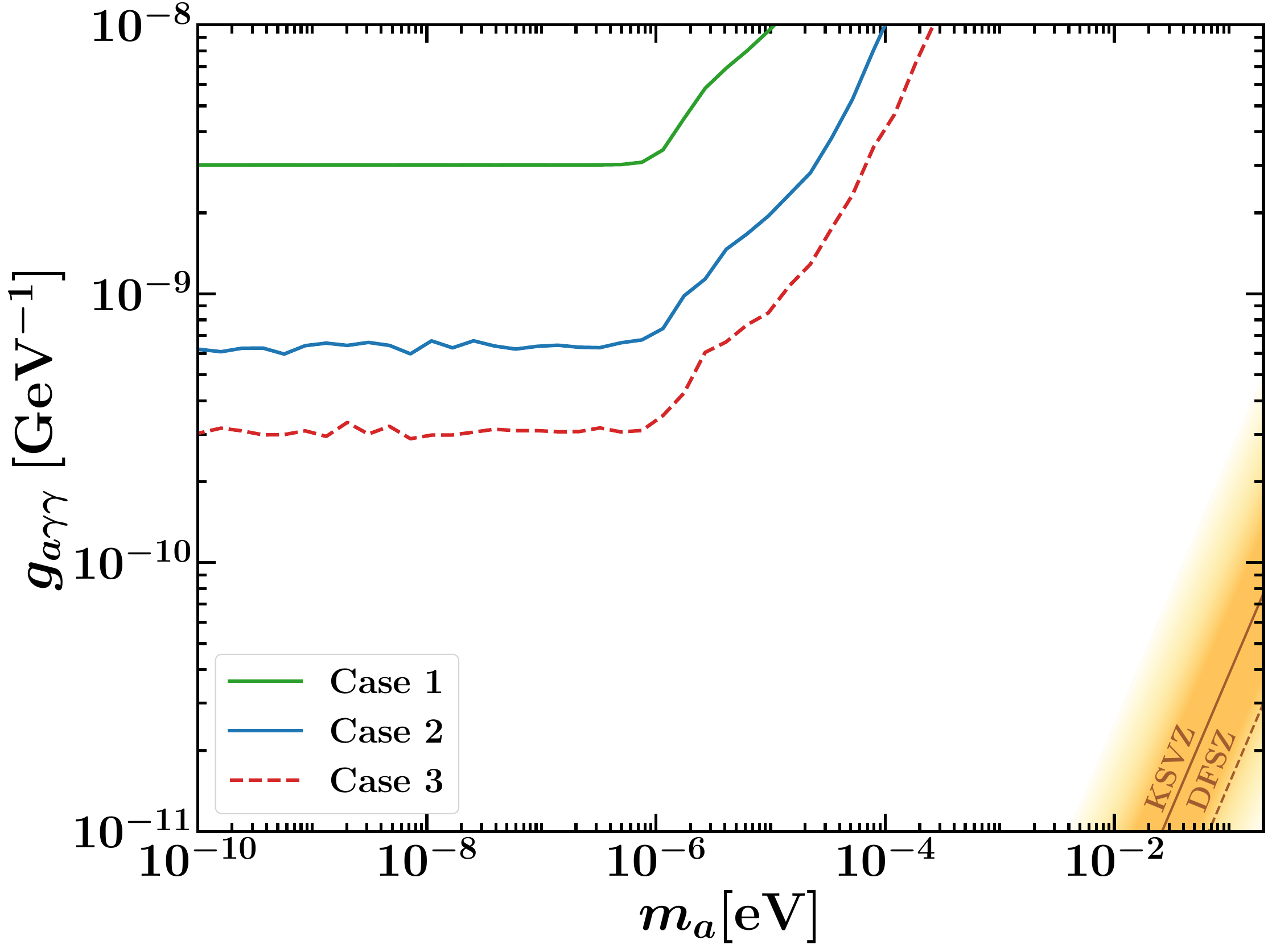}
    \caption{\footnotesize Upper limits at 95\% confidence level on the product $|g_{aN,\,\mathrm{Kr}}^{\rm eff} \times \gagg|$ (left panel) and  $\gagg$ for $g_{aN,\,\mathrm{Kr}}^{\rm eff} = 1.69 \times 10^{-6}$(right panel) as a function of the axion mass, derived from Chandrayaan-2 XSM observations. The constraint is obtained from the non-observation of an excess at the 9.4 keV $^{83}$Kr nuclear transition in the solar X-ray spectrum.}
    \label{fig:constraint_Kr}
\end{figure}

{ Finally, to derive constraints in the $m_a$--$|g_{aN}^{\rm eff}\times\gagg|$
plane, we adopt a uniform prior
$|g_{aN}^{\rm eff}\times\gagg| \in [10^{-20},\,10^{-15}]~\mathrm{GeV}^{-1}$
and construct the posterior probability density as
$\mathcal{P}(|g_{aN}^{\rm eff}\times\gagg|\,;\,\mathrm{data})$ by
multiplying the Gaussian likelihood, Eq.~\eqref{eq:likelihood}, with this
prior. The 95\% credible upper
limit on $|g_{aN}^{\rm eff}\times\gagg|$ is then obtained by integrating the
posterior probability density up to the value below which 95\% of the total posterior
probability is enclosed. This entire process is done using the \texttt{PyMultiNest} implementation of the
\texttt{MultiNest} algorithm~\cite{Feroz:2008xx}. }Repeating this procedure over
$m_a\in[10^{-12},\,10^{-3}]~\mathrm{eV}$ yields mass-dependent upper limits
on $|g_{aN}^{\rm eff}\times\gagg|$.
The constraints on $|g_{aN,\,\rm Fe}^{\rm eff}\times\gagg|$ from the
$14.4~\mathrm{keV}$ line search from $^{57}$Fe nuclear transition are shown in the left panel of
Fig.~\ref{fig:constraint_Fe}, and those on 
$|g_{aN,\,\rm Kr}^{\rm eff}\times\gagg|$ from the $9.4~\mathrm{keV}$ line search
from $^{83}$Kr nuclear transition are shown in the left panel of Fig.~\ref{fig:constraint_Kr}. 

We additionally derive independent constraints on the axion-photon coupling $\gagg$ by fixing the effective axion-nucleon couplings to $g_{aN,\,\rm Fe}^{\rm eff} = 3.6\times10^{-6}$ and $g_{aN,\,\rm Kr}^{\rm eff} = 1.69\times10^{-6}$ (the value allowed by previous searches for the $9.4$ keV line as given in~\cite{Gavrilyuk:2014mch}). Following the same procedure as before to find the 95\% credible upper limits, the resulting constraints on $\gagg$ from the $14.4~\mathrm{keV}$ $^{57}$Fe line search are displayed in the right panel of Fig.~\ref{fig:constraint_Fe}. Similarly, the corresponding limits derived from the search for the $9.4~\mathrm{keV}$ $^{83}$Kr line are shown in the right panel of Fig.~\ref{fig:constraint_Kr}.
In both panels of Fig.~\ref{fig:constraint_Fe} and Fig.~\ref{fig:constraint_Kr}, the upper limits are shown using the green solid line for conservative,  blue solid line for realistic, and red dashed line for optimistic case at 95\% C.L., respectively. 
{Note that for $m_a \lesssim 10^{-6}$ eV, constraints are essentially independent of axion mass, whereas they monotonically weaken at higher masses. This behavior directly reflects that the axion-photon conversion probability in the solar atmosphere remains constant for low masses but falls rapidly for higher masses due to loss of coherence (see Fig.~\ref{fig:osc_prob} and the discussion thereafter).}
Additionally, the summary of the 95\% C.L. upper limits on various couplings is also provided in Table~\ref{tab:CouplingValues} for $m_a \lesssim 10^{-6}\,\mathrm{eV}$.

To facilitate comparison with previous results, both panels of Fig.~\ref{fig:constraint_Fe} include existing limits on $|g_{aN,\,\rm Fe}^{\rm eff}\times\gagg|$ and on $\gagg$ (assuming $g_{aN,\,\rm Fe}^{\rm eff} = 3.6\times10^{-6}$) derived from $14.4~\mathrm{keV}$ axion searches. These prior constraints are indicated by shaded regions, where the gray region represents the CAST experiment~\cite{CAST:2009jdc}, and red represents the limits of the CUORE experiment~\cite{CUORE:2012ymr}. As evident from the figure, our conservative Case~1 limit is weaker than the CUORE bound but slightly more stringent than the CAST result. In contrast, the more realistic and optimistic background subtraction schemes (Case~2 and Case~3, respectively) produce constraints that surpass both the CAST and CUORE limits. 

To understand why our analysis yields competitive or strictly stronger coupling limits than CAST, it is instructive to compare the raw event limits. 
By treating the expected number of signal events $N_S$ as a constant across
the mass range, we find that for an exposure of
{$T_{\rm obs} = 3.72 \times 10^6\ (2.15 \times 10^6)~\mathrm{s}$},
our 95\% C.L. upper limits for Case~1, Case~2, and Case~3 are, respectively,
{
\begin{equation}
    N_{S}^{95} \lesssim 2466\,(1425),\quad 446\,(258),\quad\text{and}\quad 170\,(98).
\end{equation}
}
In contrast, the CAST experiment~\cite{CAST:2009jdc} reported a stricter raw event limit of $N_S < 32$, derived from $203$ hours ($\equiv 7.30\times 10^5$ s) of Sun tracking with an effective area $29~\mathrm{cm}^2$. 

{Naively, one might expect CAST's lower event count to translate directly into stronger constraints on the couplings. However, from Eq.~\eqref{eq:NS_def}, one can note that in reality $N_S$ is a function of axion couplings and the axion-photon conversion probability $\hat{P}(E,m_a,\gagg=1.0)$. Thus the upper limit on the coupling relies fundamentally on the axion-photon conversion probability. The CAST Phase~I data utilized for the $14.4~\mathrm{keV}$ line search was taken with evacuated magnet bores, which means the experiment relied entirely on vacuum conversion where there could be no resonance. In contrast, our search leverages conversion within the solar atmosphere (a magnetized plasma). For low axion masses, the effective photon mass in the solar atmosphere induces resonant conversion. This drastically enhances the oscillation probability, more than compensating for the XSM's smaller effective area and resulting in a stronger bound on the coupling even with our higher $N_S$ bounds.}
However this advantage is highly mass-dependent.

As seen in Fig.~\ref{fig:constraint_Fe}, for $m_a \gtrsim 10^{-5}~\mathrm{eV}$, the momentum mismatch leads to a loss of coherence in the solar atmosphere, causing the resonant oscillation probability to fall sharply. Above this threshold, our limits become considerably weaker than those of CAST, which maintains coherence over the length of its laboratory magnetic field.

Turning to the $9.4~\mathrm{keV}$ line originating from the $^{83}$Kr nuclear transition (Fig.~\ref{fig:constraint_Kr}), we found that prior constraints on the specific parameter spaces considered here ($|g_{aN,\,\rm Kr}^{\rm eff}\times\gagg|$ and $\gagg$) are not available in the literature. Consequently, our analysis represents the first dedicated search for $9.4~\mathrm{keV}$ solar axions utilizing this channel. The only existing limit related to this transition is reported by Gavrilyuk \textit{et al.}~\cite{Gavrilyuk:2014mch}, which placed an upper bound solely on the effective axion-nucleon coupling, finding $g_{aN,\,\rm Kr}^{\rm eff} < 1.69\times10^{-6}$ at 95\% C.L. for axion masses $m_a < 130~\mathrm{eV}$.

\begin{figure}
    \centering
    \includegraphics[width=0.48\linewidth]{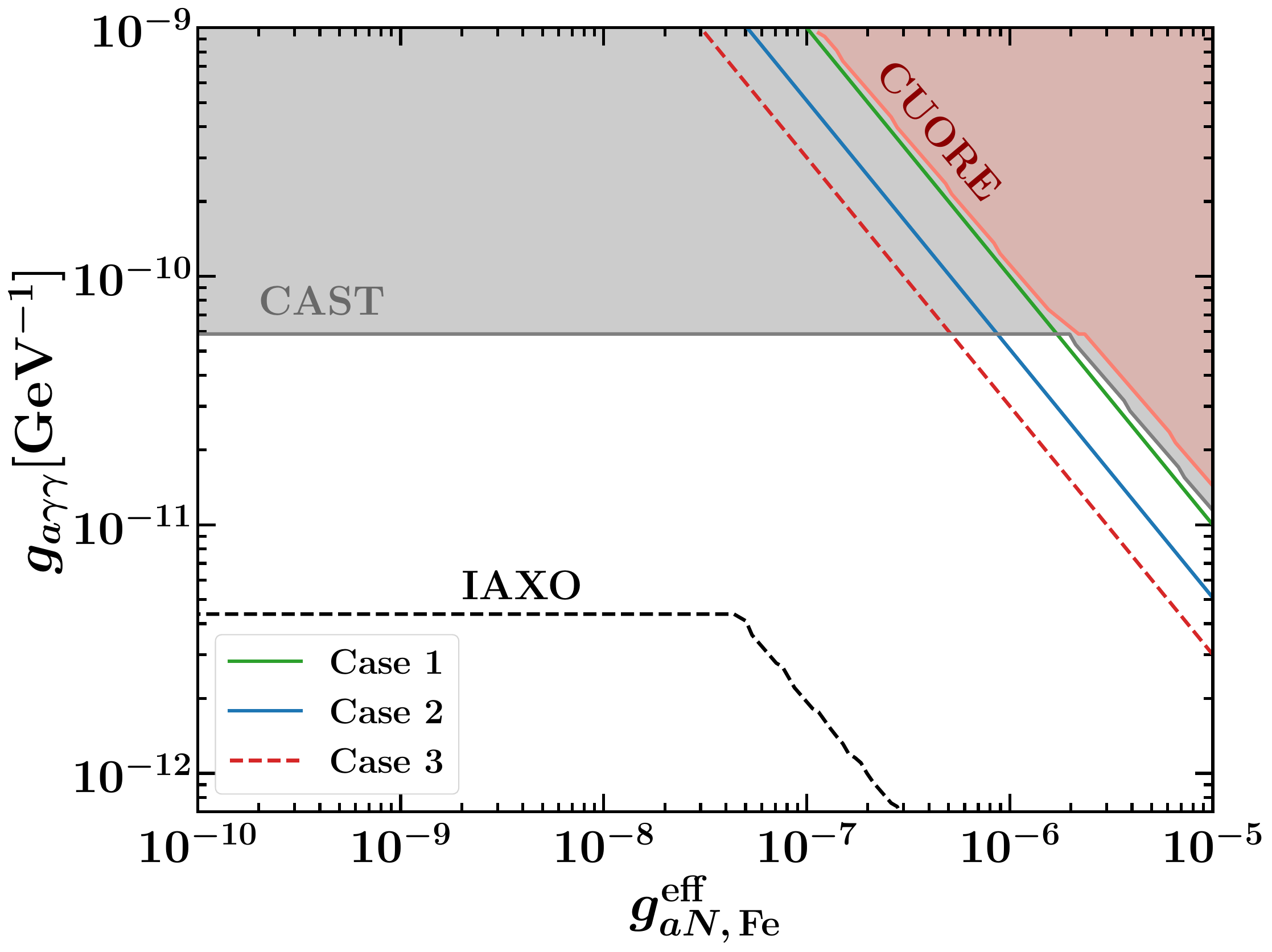}
        \includegraphics[width=0.48\linewidth]{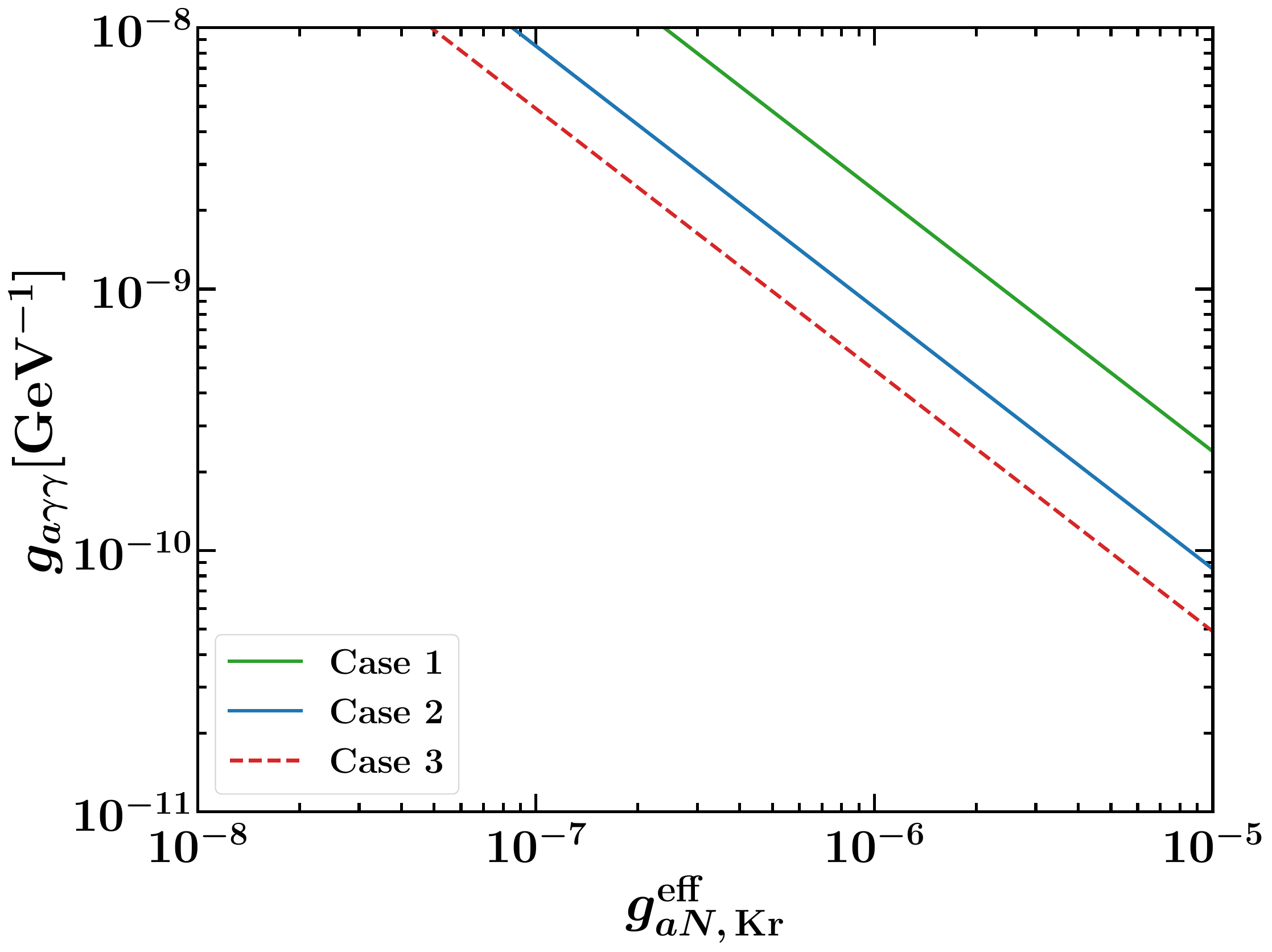}
    \caption{{ 95\% C.L. exclusion regions in the $g_{aN}^{\rm eff}$--$g_{a\gamma\gamma}$ plane (for $m_a \simeq 0$) derived from the non-observation of an excess at the 14.4~keV $^{57}$Fe (\textit{left panel}) and 9.4~keV $^{83}$Kr (\textit{right panel}) nuclear transitions in the Chandrayaan-2 XSM solar X-ray spectrum. The green solid, blue solid, and red dashed lines correspond to the conservative (Case~1), realistic (Case~2), and optimistic (Case~3) background-subtraction schemes, respectively.}}
    \label{fig:geff_vs_gagg}
\end{figure}

{We also present our results
in the $g_{aN}^{\rm eff}$--$g_{a\gamma\gamma}$ plane in
Fig.~\ref{fig:geff_vs_gagg}, for the three background subtraction schemes,
%
%
for both the $^{57}$Fe (left panel) and $^{83}$Kr (right panel) channels.
In the left panel, exclusion contours are shown alongside
existing constraints from CAST~\cite{CAST:2009jdc} (gray shaded region) and
CUORE~\cite{CUORE:2012ymr} (red shaded region), as well as the projected sensitivity of IAXO~\cite{DiLuzio:2021qct} (black dashed line), allowing a direct comparison of our limits with the existing landscape of
axion-nucleon vs axion-photon coupling constraints.
It can be seen that for $g_{aN}^{\rm
eff}\gtrsim 3\times 10^{-7}$ all three cases begin to probe parameter
space beyond the CAST exclusion,  eventually intersecting the
CUORE-excluded region toward the upper end of the $g_{aN}^{\rm eff}$
range considered.
%
%
%
For the right panel, since there are no comparable
laboratory or astrophysical constraints on $g_{a\gamma\gamma}$ currently
exist over this range of $g_{aN,\rm Kr}^{\rm eff}$, only the
XSM-derived limits for the three background-subtraction schemes are
shown. 
For $g_{aN}^{\rm eff}\sim3.6\times 10^{-6}$, the exclusion on
$g_{a\gamma\gamma}$ recovers the reference limits already shown in
Figs.~\ref{fig:constraint_Fe} and~\ref{fig:constraint_Kr}.}

{Before concluding this section, we would like to mention that our derived limits are affected by the uncertainties in the solar elemental abundance, in the Chandrayaan-2
background modeling, and in the solar magnetic field, all of which has been discussed in Appendix~\ref{app:Bfield_systematic}.}

\section{Results and Discussion}
\label{sec:results}

In this work, we have investigated the production of monoenergetic solar axions
via the M1 nuclear de-excitation of $^{57}$Fe and $^{83}$Kr, emitting at
$14.4\,\mathrm{keV}$ and $9.4\,\mathrm{keV}$, respectively. These axions
can subsequently convert into X-ray photons in the solar magnetic field
via the inverse Primakoff process, yielding a sharp spectral line
observable by X-ray telescopes. Using solar soft X-ray data from
Chandrayaan-2 XSM, we have derived constraints on the product of the
effective axion-nucleon and axion-photon couplings,
$|g_{aN}^{\rm eff} \times \gagg|$ and{the axion-photon coupling} $\gagg$ alone for a fixed $g_{aN}^{\rm eff}$, across three background
subtraction schemes, which we refer to as conservative, realistic,
and optimistic. For a summary, see the second and third columns of Table~\ref{tab:CouplingValues} for $^{57}$Fe.
The conservative bound is comparable to existing constraints from CAST and CUORE, while the realistic and optimistic cases surpass them. 
{Despite XSM's smaller effective area compared to CAST, we find that the longer solar observation time and enhanced conversion probability arising from resonant conversion in the solar atmosphere yield stronger limits in both the realistic and optimistic cases.}

For $^{83}$Kr (see 4th and 5th columns Table~\ref{tab:CouplingValues}), these bounds are weaker than those
from $^{57}$Fe by more than an order of magnitude, a direct consequence
of the lower solar abundance of krypton, which suppresses the
$^{83}$Kr axion flux by nearly three orders of magnitude
as discussed in Sec.~\ref{sec:solar_axion_production}.
The effective axion-nucleon couplings for the two isotopes,
$g_{aN,\,\rm Fe}^{\rm eff} = -1.19\,g_{0aN} + g_{3aN}$ and
$g_{aN,\,\rm Kr}^{\rm eff} = -g_{0aN} + g_{3aN}$, are numerically
similar but slightly different linear combinations of the
isoscalar and isovector nucleon couplings. Taken together, the two
lines therefore offer complementary sensitivity to the axion-nucleon
parameter space.

\begin{table}
\centering
\large
\begin{tabular}{|c|cc|cc|}
\hline
\multirow{2}{*}{\bf Case} & \multicolumn{2}{c|}{$^{\boldsymbol{57}}$\textbf{Fe}}                                 & \multicolumn{2}{c|}{$^{\boldsymbol{83}}$\textbf{Kr}}                                  \\ \cline{2-5} 
 &
  \multicolumn{1}{c|}{\begin{tabular}[c]{@{}c@{}}$|g_{aN,\,\rm Fe}^{\rm eff} \times \gagg| \lesssim$ \\ {[}GeV$^{-1}${]}\end{tabular}} &
  \begin{tabular}[c]{@{}c@{}}$\gagg \lesssim$ \\ {[}GeV$^{-1}${]}\end{tabular} &
  \multicolumn{1}{c|}{\begin{tabular}[c]{@{}c@{}}$|g_{aN,\,\rm Kr}^{\rm eff} \times \gagg| \lesssim$ \\ {[}GeV$^{-1}${]}\end{tabular}} &
  \begin{tabular}[c]{@{}c@{}}$\gagg \lesssim$ \\ {[}GeV$^{-1}${]}\end{tabular} \\ \hline
\textbf{Conservative}          & \multicolumn{1}{c|}{$1.2\times10^{-16}$} & $3.5\times10^{-11}$ & \multicolumn{1}{c|}{$2.5\times10^{-15}$} & $ 3.0 \times 10^{-9}$  \\ \hline
\textbf{Realistic}             & \multicolumn{1}{c|}{$5.0\times10^{-17}$}   & $1.5\times10^{-11}$ & \multicolumn{1}{c|}{$8.5\times10^{-16}$} & $ 6.0 \times 10^{-10}$ \\ \hline
\textbf{Optimistic}            & \multicolumn{1}{c|}{$3.0\times10^{-17}$}   & $8.0\times10^{-12}$   & \multicolumn{1}{c|}{$5.0\times10^{-16}$}   & $ 3.0 \times 10^{-10}$ \\ \hline
\end{tabular}
\caption{\footnotesize Summary of the 95\% C.L. upper limits obtained on $|g_{aN,\,\mathrm{Fe}}^{\rm eff} \times \gagg|$ and $\gagg$ from the $14.4$ keV line search from $^{57}$Fe transition and $|g_{aN,\,\mathrm{Kr}}^{\rm eff} \times \gagg|$ and $\gagg$ the $9.4$ keV line search from $^{83}$Kr for $m_a \lesssim 10^{-6}\,\mathrm{eV}$.}
 \label{tab:CouplingValues}
\end{table}

{Note that the terrestrial experiments like CAST~\cite{CAST:2009jdc} and CUORE~\cite{CUORE:2012ymr}  have previously investigated monochromatic solar axion flux.
In contrast to these, we use the XSM aboard Chandrayaan-2 to search for a characteristic narrow spectral feature around 14.4~keV and 9.4 keV in space-based solar X-ray observations.
This provides a complementary probe of the axion-nucleon and axion-photon couplings using an existing solar X-ray dataset rather than 
conventional laboratory searches.
Our analysis therefore establishes a novel space-based and astrophysical strategy for probing nuclear transition solar axions.
}


\section*{Acknowledgments}

The authors thank Maurizio Giannotti for his insightful comments, which helped to improve the manuscript significantly. The authors gratefully acknowledge Santosh Vadawale, N. P. S. Mithun, and B. S. Bharath Saiguhan for providing the reduced Chandrayaan-2 dataset utilized in this analysis. The authors are also grateful to the organizers of the Workshop in High Energy Physics (WHEPP) 2025 at Indian Institute of Technology Hyderabad, where this work was initiated. TK would also like to acknowledge support in the form of a Senior Research Fellowship from the Council of Scientific \& Industrial Research (CSIR), Government of India.

{
\begin{appendices}
\section{Discussion on Systematic uncertainties}
\label{app:Bfield_systematic}

{Here, we comment on the different sources of
uncertainty that can affect our presented limits, namely the uncertainty
in the solar elemental abundance, the uncertainty in the Chandrayaan-2
background modeling, and the uncertainty in the solar magnetic field.

{The uncertainty
associated with the solar elemental abundance affects our derived limits. }
Since the axion flux normalization $\Phi_0^{\rm Fe/Kr}$ in Eq.~\eqref{eq:normalization_flux} scales linearly with
the elemental abundance, $10^{\epsilon_X}$ (see
Eq.~\eqref{eq:flux}), and the derived $95\%$ C.L. upper limits on the coupling scale as
$\Phi_0^{-1/2} \propto 10^{-\epsilon_X/2}$,
any uncertainty in the adopted solar abundance directly propagates to an
uncertainty in our derived coupling limits. Using the uncertainties in the solar photospheric Fe and Kr abundances
$\log\epsilon_{\rm Fe} = 7.46\pm0.04$ and $\log\epsilon_{\rm Kr} =
3.12\pm0.10$ from Ref.~\cite{2021A&A...653A.141A}, and with the above scaling of the limits with the abundance, the corresponding
$1\sigma$ uncertainties on our derived coupling limits are approximately $\pm\, 4.5\%$ for the Fe channel and $\pm\,11.5\%$ for the Kr channel.
}

{
{The uncertainty associated with the background modeling is covered by the three  cases we have considered in our analysis, see Sec.~\ref{sec:xsm_details}.}
Case~1--Case~3 explicitly show the degree to which the derived limits
depends on background modeling. The resulting limits are summarized in
Table~\ref{tab:CouplingValues}, and the spread between the three cases quantifies the
systematic uncertainty associated with background subtraction. 
We report
Case~2 as our baseline (realistic) constraint throughout this work, with
Case~1 as the conservative bound and Case~3 as an idealized target limit.

We note that Case~3 indicates the best sensitivity one can expect
from this dataset, limited only by counting statistics under an ideal
background subtraction. Such an ideal subtraction would require a
comprehensive account of all contributions to the detector background  the cosmic X-ray background (CXB), the charged particle background of
cosmic and solar origin, and secondary emission from their interaction with
the spacecraft structure together with the variable occultation of the
CXB and particle background within the FOV by the Moon, and the presence of
variable bright astrophysical sources within the wide XSM FOV. While
Case~2 accounts for most of these effects, some components, such as
spacecraft secondary emission and the time variability of fainter sources
within the FOV, are difficult to model and are not fully captured, and
addressing them would be necessary to approach the ideal Case~3. We note
that the empirical background model underlying Case~3 is a smooth power law
fit to the broadband spectrum, with no narrow spectral components. Since the
expected axion-induced signal is a narrow line at a fixed, known energy, it
is not expected to be systematically absorbed by this background model.
Case~3 should therefore be interpreted as a target sensitivity motivating
further work on accurate background modeling, rather than as our primary
result.
}


\begin{figure}
    \centering
    \includegraphics[width=0.55\linewidth]{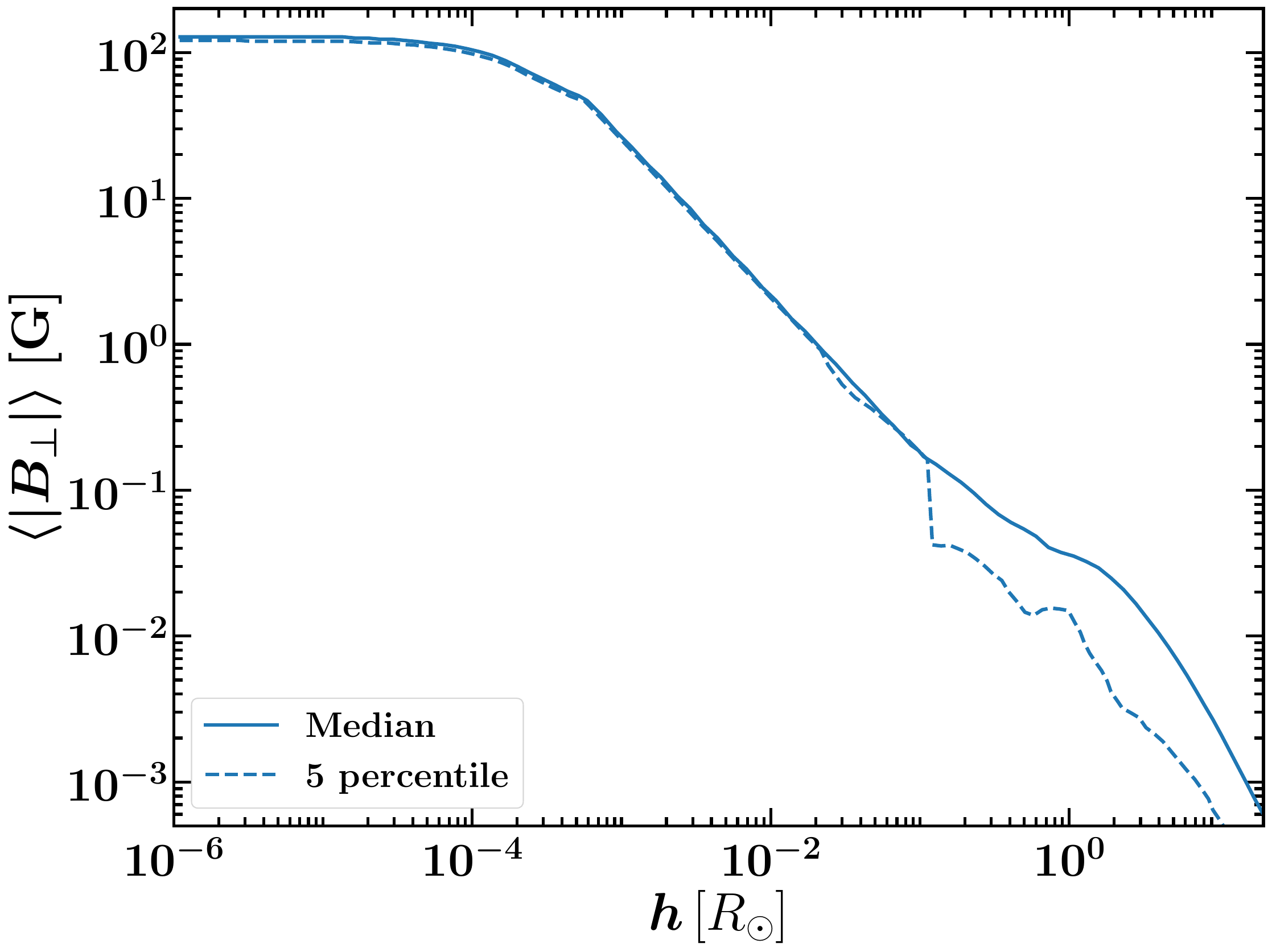}
   \caption{Median (solid) and 5th-percentile (dashed) coronal magnetic field
profiles $\langle|B_\perp|\rangle$ as a function of altitude, adapted from Ref.~\cite{Ruz:2024gkl}.}
    \label{fig:solar_magnetic_field}
\end{figure}

Finally, there is the uncertainty in the solar magnetic field modeling,
particularly the magnetic field in the solar corona. The axion-photon conversion probability $P_{a\gamma}$ used throughout this work is computed using the coronal magnetic field model of
Ref.~\cite{Ruz:2024gkl}, based on a Predictive Science Inc. (PSI) MHD
simulation of the solar corona during the total solar eclipse of July 2,
2019. In the coherent (low-mass) regime, $P_{a\gamma}\propto B^2$, so the derived coupling limits depend sensitively on this field model.
The PSI simulation is a snapshot at a single epoch, while our
Chandrayaan-2 observing seasons span September - November 2019 and
February - May 2020 (Sec.~\ref{sec:xsm_details}). Ref.~\cite{Ruz:2024gkl}
cross validated the same field model against a same day
Potential Field Source Surface (PFSS) reconstruction for a NuSTAR solar
observation on February 21, 2020 a date lying within our second
observing season and found good agreement. In our case, the first observing season of Chandrayaan-2 lies closer to the
eclipse epoch (2 - 5 months) than this validated interval, while the later part of the second season extends slightly beyond it (up to
$\sim$ 10 - 11 months). A day resolved coronal field model spanning  full observing window of Chandrayaan-2 is not available, so we adopt the same fiducial (median) profile as Ref.~\cite{Ruz:2024gkl}.

\begin{figure}[htbp]
    \centering
    \includegraphics[width=0.45\textwidth]{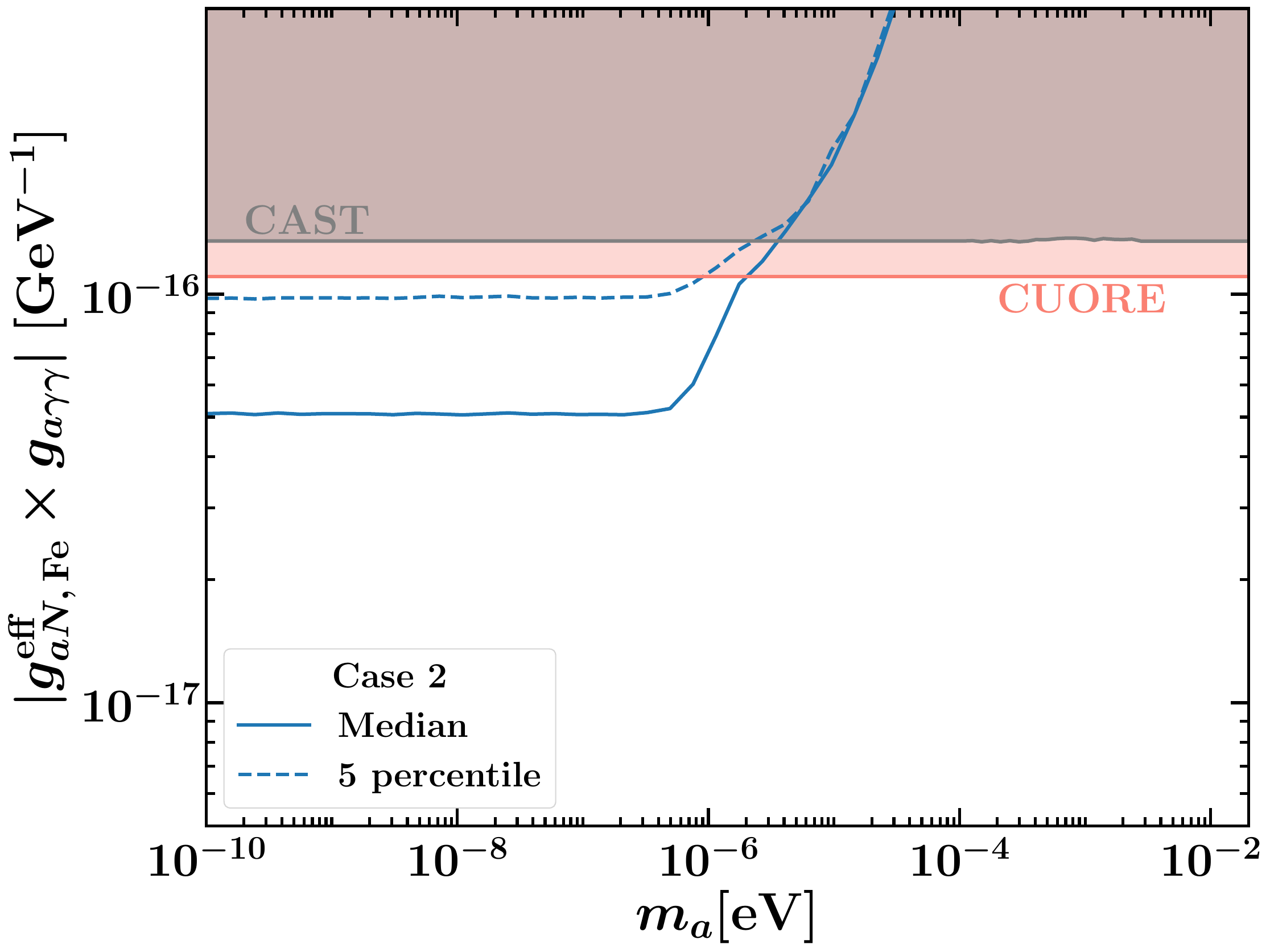}
    \includegraphics[width=0.45\textwidth]{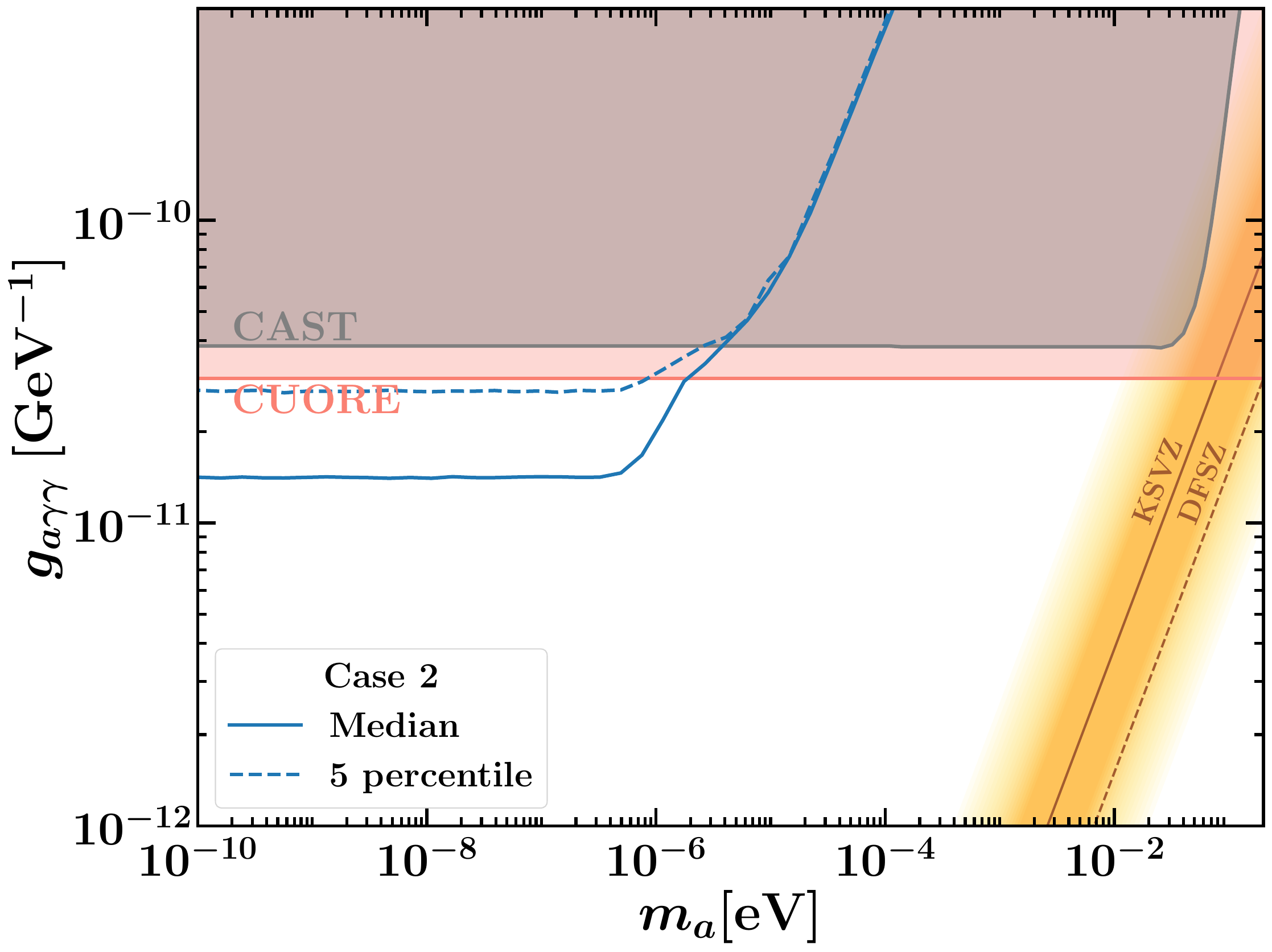}
    \caption{Effect of the coronal magnetic field uncertainty on the
    Case~2 (realistic) 95\% C.L. limit for the $^{57}$Fe channel, on
    $|g_{aN,\rm Fe}^{\rm eff}\times g_{a\gamma\gamma}|$
    (\textit{\textbf{left panel}}) and on $g_{a\gamma\gamma}$
    (\textit{\textbf{right panel}}, for
    $g_{aN,\rm Fe}^{\rm eff}=3.6\times10^{-6}$). The solid line uses the
    fiducial (median) coronal field profile; the dashed line uses the
    5th-percentile (weak-field) profile from Ref.~\cite{Ruz:2024gkl}.}
    \label{fig:Bfield_systematic}
\end{figure}

They constructed an ensemble of 120 coronal field
profiles by azimuthally rotating the PSI simulation in $3^\circ$
increments, reporting the median as fiducial along with the 5th, 10th, and
15th percentiles as weaker-field alternatives. {In this work,}  replacing the median with
the 5th-percentile profile, and keeping all other aspects of the analysis
unchanged, {we find for  Case~2 that the limits weakens at 95\% C.L. limit for the
$^{57}$Fe channel from $|g_{aN,\rm Fe}^{\rm eff}\times g_{a\gamma\gamma}|
\lesssim 5.0\times10^{-17}$ to $\sim1.0\times10^{-16}~\mathrm{GeV}^{-1}$} {(see the left-panel of Fig.~\ref{fig:Bfield_systematic})}. The corresponding limit on $g_{a\gamma\gamma}$ (for
$g_{aN,\rm Fe}^{\rm eff}=3.6\times10^{-6}$) changes from $1.41\times10^{-11}$ to
$2.73\times10^{-11}~\mathrm{GeV}^{-1}$, in both cases a degradation by a
factor of $\sim2$ {(see the right-panel of Fig.~\ref{fig:Bfield_systematic})}. This is consistent with our expectation,
given the roughly
factor of two difference in field strength between the two profiles, and
we note it as representative of the systematic uncertainty on our bound
from the coronal field model. However we find that even with the weaker field strength our realistic limit is stronger than those obtained from either CAST or CUORE for $m_a \lesssim 10^{-6}$ eV. We also note that given the magnetic field remains same, we also expect to get a similar effect for the limits obtained from the $^{83}$Kr channel

\end{appendices}
}

\bibliographystyle{JHEP}
\bibliography{refs}

\end{document}